\documentclass[a4paper,11pt]{article}

\usepackage[margin=1in]{geometry} 

\usepackage{amsmath}
\usepackage{amsthm}
\usepackage{amssymb}
\usepackage{lineno}

\usepackage{placeins}

\usepackage{etoolbox}

\usepackage{graphicx}
\usepackage{bm}
\usepackage{lineno}
\usepackage{caption}
\usepackage{subfigure}	

\usepackage{booktabs}

\usepackage[breaklinks]{hyperref}

\usepackage{multirow}

\usepackage{comment}
\usepackage{xargs}                      

\usepackage[utf8]{inputenc}

\hypersetup{
	unicode,
	colorlinks=true,
	linkcolor=blue,
	filecolor=magenta,      
	citecolor=blue,
	urlcolor=cyan,
	pdfauthor={ M. A. Mendez},
	pdftitle={Linear and Nonlinear Dimensionality Reduction from Fluid Mechanics to Machine Learning},
	pdfsubject={Linear and Nonlinear Dimensionality Reduction from Fluid Mechanics to Machine Learning},
	pdfkeywords={article, template, simple},
	pdfproducer={LaTeX},
	pdfcreator={pdflatex}
	colorlinks=true,
	citecolor=blue,
	linkcolor=blue,
	filecolor=magenta,      
	urlcolor=blue,
}

\usepackage[most]{tcolorbox}
\usepackage{booktabs}
\usepackage{amsmath}
\usepackage{wrapfig}

\usepackage{mathtools}
\tcbset{enhanced,colback=cyan!2!white,colframe=cyan!90!white,coltitle=blue!20!black,fonttitle=\bfseries}

\usepackage{listings}
\usepackage{hyperref}
\hypersetup{
	colorlinks=true,
	linkcolor=blue,
	filecolor=magenta,      
	citecolor=blue,
	urlcolor=cyan,
}

\usepackage[sort&compress,authoryear,round]{natbib}
\usepackage{algorithm, algpseudocode} 
\usepackage{mathrsfs} 

\title{Linear and Nonlinear Dimensionality Reduction from Fluid Mechanics to Machine Learning}
\author{Miguel A. Mendez}

\date{
	von Karman Institute for Fluid Dynamics, Sint-Genesius-Rode, Belgium %
}

\begin{document}
	\maketitle
	
	\begin{abstract}

Dimensionality reduction is the essence of many data processing problems, including filtering, data compression, reduced-order modeling and pattern analysis. While traditionally tackled using linear tools in the fluid dynamics community, nonlinear tools from machine learning are becoming increasingly popular. This article, halfway between a review and a tutorial, introduces a general framework for linear and nonlinear dimensionality reduction techniques. Differences and links between autoencoders and manifold learning methods are highlighted, and popular nonlinear techniques such as kernel Principal Component Analysis (kPCA), isometric feature learning (ISOMAPs) and Locally Linear Embedding (LLE) are placed in this framework. These algorithms are benchmarked in three classic problems: 1) filtering, 2) identification of oscillatory patterns, and 3) data compression. Their performances are compared against the traditional Proper Orthogonal Decomposition (POD) to provide a perspective on their diffusion in fluid dynamics.

		\vspace{7mm}
		
	\end{abstract}

	
	
	\clearpage
	\tableofcontents
	\clearpage

	\section{Introduction} \label{sec:1}
	
	Dimensionality reduction is an essential branch of machine learning and data engineering, concerned with identifying ways to encode high dimensional data in a low dimensional representation (see \cite{Alpaydin2020-ky,Bishop2011,Murphy2012}). Many of its methods are widespread across different disciplines under different names. 
	
	In fluid dynamics, the popularity of dimensionality reduction can be traced back to the introduction of the Proper Orthogonal Decomposition (POD) as a tool to identify (and define) coherent structures in turbulent flows \citep{Lumley1,Lumley1997}. The idea that a seemingly chaotic dynamic could be pictured as a linear combination of coherent structures has fueled the interest in the ``dynamical system perspective'' of turbulent flows (see \cite{Berkooz1993,Holmes1996}) and the development of model order reduction techniques based on Galerkin projection (see \cite{CORDIER,Benner2015,Shady,Benner2020-ta} for exhaustive reviews).
	
	The diffusion of the POD in both experimental and numerical fluid dynamics is largely due to the works of \cite{Siro1,Siro2,Siro3} and \cite{Aubry,Aubry1991} who have proposed efficient algorithms to compute the POD and analyzed their link to the original formulation by \cite{Lumley1} (see also \cite{George2016}). With the continuous development of experimental fluid mechanics-- and in particular Particle Image Velocimetry (PIV)--, the use of POD to educe coherent structures from experimental data has boomed in the last two decades (see for example \cite{Sullivan1996,CITRINITI2000,gordeyev_thomas_2000,Bi2003,Semeraro2012,Pollard2017,Mendez2020,Mendez2018}). Focusing on the literature in experimental fluid dynamics, the POD has also been extensively used as a filter, for example to remove outliers in PIV measurement \citep{Raiola_2015,Higham_2016} or to pre-process velocity fields for pressure integration \citep{Charonko2010}, to pre-process images \citep{MENDEZ2017181}, to fill `gaps' in experimental data \citep{Saini2016} to construct efficient regressors and interpolators \citep{Ratz2022,Casa2013,Karri2009,Bouhoubeiny2009}, to validate numerical simulations \citep{Kriegseis_2010} or to build estimators of quasi-periodic flows \citep{Bourgeois2013,Loiseau2018}. 
	Moreover, the POD has been used to enhance adaptive least square problems (see \cite{Yao2017}), fault diagnostics \citep{Shen2022} or optimal pressure placement \citep{Castillo2020}. 
	
	In the last decade, many variants and hybrid formulations of the POD have been developed (see \cite{sieber_paschereit_oberleithner_2016,towne_schmidt_colonius_2018,Mendez2019}) alongside alternative data-driven decompositions. Among these, the most popular alternative is the Dynamic Mode Decomposition (DMD) proposed by \cite{Schmid} and \cite{Rowley2}, also existing in many variants (see \cite{Kutz2014,Jovanovic2014,Kutz2016,rec_DMD,Clainche2017}). The DMD can be seen as a variant of the Principal Oscillation Pattern (POP), and Linear Inverse Method (LIM) approaches diffused in climatology \citep{Hasselmann1988,Storch1990,Penland1996} in the late '80s. Although differing slightly in their algorithmic implementation, all DMD formulations fit a linear dynamical system to the dataset to identify the dominant oscillatory patterns.

	More broadly, the study of ways of decomposing a flow as a linear combination of ``coherent patterns'', referred to as ``modes'' has evolved into a well-established branch of data processing, often referred to as data-driven modal analysis (see \cite{Taira,Taira2020}). Within the machine learning literature, all modal decompositions fall in the category of \emph{linear} dimensionality techniques. Although fundamental both conceptually and in practice, these are only a fraction of dimensionality reduction methods routinely used in data science (see \cite{Velliangiri2019,Ghojogh2019a,Ayesha2020}). Most of the recent development in the field focused on \emph{nonlinear} methods for dimensionality reduction. These, somewhat surprisingly, are currently relatively unexplored in fluid dynamics. A recent review on these methods is provided by \cite{Csala2022}, who focused on their application to numerical data.
	
	Among (nonlinear) dimensionality reduction techniques, one can identify two main classes of methods: (1) autoencoders and (2) manifold learning methods. Autoencoders seek a compressed representation of the data while preserving as much information as possible. The POD -- known as Principal Component Analysis (PCA) in the machine learning literature \citep{Ghojogh2019}-- is a linear autoencoder \citep{Baldi1989,Milano2002}. Manifold learning methods are not concerned with the loss of information in the reconstruction but seek low dimensional representation that preserves as much as possible some measure of similarity \citep{Zheng2009}. The difference between these two classes of methods is subtle (some autoencoders can be used for manifold learning and vice-versa) but essential: most algorithms for manifold learning lead to compressed representations that do not admit `an inverse', that is a mapping back to the original space.
	
	Among the autoencoders, the most classic nonlinear approaches are Artificial Neural Networks (ANN) autoencoders \citep{GoodBengCour16,Alain2012} and kernel PCA \citep{Schoelkopf1997,Ghojogh2019}. While ANN are becoming increasingly popular in fluid dynamics \citep{Agostini2020,Pawar,Fukami2021,Hamidreza,Fukami2020,guastoni,Singh_2020}, the kPCA has not yet been exploited in the field. Among the manifold learning techniques, the most popular ones are arguably the Locally Linear Embedding (LLE) \cite{LLE_Ref,Saul2001AnIT,Ghojogh2020a} and ISOMAPs \citep{Tenenbaum2000,Ghojogh2020}. These are now entering the fluid dynamics  comunity as alternatives to the POD for reduced order modeling \citep{Chen2022F} and for finding compressed representation of fluid flows \citep{Ehlert2019,Farzamnik2022,Tauro2014}.
	
	The scope of this work is to provide a tutorial and a comparative analysis of kPCA, LLE and ISOMAP, benchmarking these relatively `new' methods against the  POD for three applications of great interest to the experimental fluid dynamicist: 1) filtering, 2) identification of oscillatory patterns, and 3) data compression. 
	
	The rest of the article is structured as follows. Section \ref{sec:2} introduces the problem set and the mathematical framework. Section \ref{sec:3} analyzes the connection between the POD and the kPCA, LLE and ISOMAP. Following the footsteps of the excellent tutorial by \cite{Ghojogh2020}, this is done by building bridges with the most elementary manifold learning methods known as Multi-dimensional scaling (MDS, see \cite{Torgerson1952}). Section \ref{sec:4} introduces the three selected test cases. The first is an image filtering problem: the background removal from PIV images. The second is a manifold learning problem, namely identifying oscillatory patterns in a vortex shedding problem. The third is a data compression problem concerned with analysing a flow configuration that does not exhibit a clear low dimensional representation. Results are presented in Section \ref{sec:5} while conclusions and perspectives are summarized in Section \ref{sec:6}.

	\section{The Problem Set}\label{sec:2}

Our starting point is a dataset rearranged in the form of a \emph{snapshot matrix} $\bm{D}\in\mathbb{R}^{n_s\times n_t}$. This contains $n_t$ snapshots (columns) collecting $n_s$ data points ech. Using a Python-like notation, we denote as $\bm{d}_k=\bm{D}[:,k]\in \mathbb{R}^{n_s}$ the k-th column (snapshot) in $\bm{D}$. The reshaping into a column vector is carried out regardless of the nature of the dataset.

Considering for example a planar PIV measurement, providing a velocity field $\bm{u}(x_i,y_j,t_k)=(u(x_i,y_j,t_k),v(x_i,y_j,t_k))^T$ over a regular grid $\mathbf{x}=(x_i,y_j)$ with $i=1,\dots n_x$ and $j=1,\dots n_y$. A snapshot vector can be obtained by flattening the two scalar fields $u,v$ into vectors and stucking these one below the other to have $n_s=2 n_x n_y$. If the dataset contains a video sequence, then each snapshot is obtained by reshaping the images.

In this article, we assume that data is available in a Cartesian and uniform grid with spacing $\Delta x$ in both directions. Then, in a first approximation one can assume that the data is constant within an area $\Delta x ^2$ (piece-wise constant interpolation). Denoting as $\Omega$ the domain (area) over which the velocity was sampled, the $L_2$ norm in $\Omega$ can then be linked to the $l_2$ norm in $\mathbf{R}^{n_s}$ (see \cite{Mendez2019}).  Similarly, assuming that the snapshots have been collected at uniform time intervals $\Delta t$, so that $t_k=k\Delta t$, with $k=0,\dots n_t-1$ and $f_s=1/\Delta t$ is the sampling frequency, the $L_2$ norm in the time interval of duration $T=n_t \Delta t$ can be linked to the $l_2$ norm in $\mathbb{R}^{n_t}$.

We here focus on dimensionality reduction technique in the \emph{space} domain, i.e. along the columns of $\mathbf{D}$. Therefore, we seek to map our dataset $\bm{D}\in\mathbb{R}^{n_s\times n_t}$ into a compressed representation $\bm{Z}\in \mathbb{R}^{n_s\times n_t}$. The column $\bm{z}_k=\bm{Z}[:,k]\in\mathbb{R}^{n_r}$ is thus the reduced representation of the snapshot $\bm{d}_k\in\mathbb{R}^{n_s}$, with $n_r\ll n_s$.

In autoencoders, the goal is to find the reduced representation that preserves most of the relevant information. We thus define two mappings. The first, denoted as \emph{encoder} $\mathcal{E}$, deals with the compression $\mathcal{E}:\mathbb{R}^{n_s}\rightarrow \mathbb{R}^{n_r}$, i.e. $\bm{Z}=\mathcal{E}(\bm{D})$. The second, which we denote as \emph{decoder} $\mathcal{D}$, deals with the reverse process $\mathcal{D}:\mathbb{R}^{n_r}\rightarrow \mathbb{R}^{n_s}$, i.e. $\tilde{\bm{D}}=\mathcal{D}(\bm{Z})$. The tilde here denotes ``approximation of'', because the mapping $\mathbb{R}^{n_s}\rightarrow \mathbb{R}^{n_r}\rightarrow \mathbb{R}^{n_s}$, denoted as \emph{autoencoder} $\mathcal{A}=\mathcal{D}(\mathcal{E}())$, is lossy. Preserving essential information means having $\tilde{\bm{D}}\approx \bm{D}$.

In manifold learning, there is usually no decoder. The goal is to find the compressed representation that preserves (as much as possible) some measure of pair-wise similarity in the original space and in the reduced space. That is, defining as $\mathcal{S}_{d}(,):\mathbb{R}^{n_s}\times\mathbb{R}^{n_s}\rightarrow \mathbb{R}$ a similarity measure in $\mathbb{R}^{n_s}$ and  $\mathcal{S}_{z}(,):\mathbb{R}^{n_r}\times\mathbb{R}^{n_r}\rightarrow \mathbb{R}$ a similarity measure in $\mathbb{R}^{n_r}$, preserving similarity means having $\mathcal{S}_{d}(\bm{d}_i,\bm{d}_j)\approx \mathcal{S}_{z}(\bm{z}_i,\bm{z}_j)$ for all $i,j$.
	
\section{Autoencoders and Manifold Learning}\label{sec:3}	

		\subsection{From POD (PCA) to MDS }\label{sec:3p1}

We begin by analyzing the link between classic POD and Multidimensional Scaling (MS). This link connects methods from modal decomposition to methods from manifold learning. Assuming uniform sampling in space and time, as mentioned in the previous section, the POD is equivalent to the Principal Component Analysis (PCA).


The POD (PCA) is a linear autoencoder. The linear encoder $\mathcal{E}:\mathbb{R}^{n_s}\rightarrow \mathbb{R}^{n_r}$ is a projection onto an orthogonal\footnote{Note that orthogonality is not a constraint but a convinient choice for an autoencoder aiming at preserving information. The most popular non-orthogonal autoencoders is the DMD.} basis of $n_r$ vectors, that is $\bm{Z}=\bm{\Phi}^T \bm{D}$, with $\bm{\Phi}\in\mathbb{R}^{n_s\times n_r}$ the matrix collecting the basis elements along the columns. These basis elements, denoted as $\bm{\phi}_r=\bm{\Phi}[:,r]$, are the coherent structures (or fields) in the data and the columns $\bm{z}_k=\bm{Z}[:,k]\in\mathbb{R}^{n_r}$ are the reduced representation of the snapshot $\bm{d}_k=\bm{D}[:,k]\in\mathbb{R}^{n_s}$ with respect to that basis. The decoder is a linear combination of these structures, that is $\tilde{\bm{D}}=\bm{\Phi} \bm {Z}$. The autoencoder is thus $\tilde{\bm{D}}=\bm{\Phi} \bm{\Phi}^T \bm {D}$, and all mappings are given once the basis matrix $\bm{\Phi}$ is given. 

In classic POD (PCA), this matrix is the one that minimizes 

\begin{equation}
	\label{COST}
	J(\bm{\Phi})=||\bm{D}-\tilde{\bm{D}}||^2_2=||\bm{D}-\bm{\Phi}\bm{ \Phi}^T \bm{D}||^2_2\,,
\end{equation}  where $||\bullet||_2$ is the induced $l_2$ norm.

To hope for a unique solution, we impose that these vectors have unitary length, hence $\bm{\Phi}^T\bm{\Phi}=\bm{I}$, with $\bm{I}$ the identity matrix of appropriate size. The minimization of $J(\bm{\Phi})$, under the orthonormality constraints, leads to the well-known eigenvalue problem (see \cite{Bishop2011})

\begin{equation}
	\label{EIG_POD}
	\bigl(\bm{D} \bm{D}^T \bigr) \bm{\Phi}_{\mathcal{P}}= \bm{\Phi}_{\mathcal{P}} \bm{\Lambda}_{\mathcal{P}}\,,
\end{equation}  with $\bm{\Lambda}_{\mathcal{P}}=\mbox{diag}(\lambda_{\mathcal{P}1},\lambda_{\mathcal{P}2}\dots \lambda_{\mathcal{P}n_r})$ the diagonal matrix collecting the non-negative eigenvalues $\lambda_r$ associated to the basis element $\phi_{\mathcal{P}r}$. A unique decomposition can be found if all eigenvalues are distinct (which is not the case in uncorrelated noise). In what follows, we use the subscript $\mathcal{P}$ to distinguish the PCA/POD bases from the others and define $\bm{C}:=\bm{D} \bm{D}^T$.

In fluid dynamics, $\bm{C}$ is the spatial correlation matrix and solving for $\bm{\Phi}$ via \eqref{EIG_POD} is known as `classic POD' or `space-only' POD \citep{George2016,towne_schmidt_colonius_2018}. 
In machine learning and statistics, $\bm{C}$ is the covariance matrix, and the classic POD is known as `classic' PCA \citep{Ghojogh2019PCA}. An alternative formulation considers the encoding (projection) along the rows rather than columns (i.e. in space rather than time in our settings).

To analyze their link, let us factorize the reduced representation as $\bm{Z}_{\mathcal{P}}=\bm{\Sigma}_{\mathcal{P}}\bm{\Psi}_{\mathcal{P}}^T$, with $\bm{\Sigma}_{\mathcal{P}}=\mbox{diag}(\sigma_{\mathcal{P}1},\dots\sigma_{\mathcal{P}n_r})\in\mathbb{R}^{n_R\times n_r}$ a diagonal matrix containing the $l_2$ norms of the rows $\bm{Z}_{\mathcal{P}}$, i.e. $\sigma_{\mathcal{P}r}=||\bm{Z}_{\mathcal{P}}[r,:]||_2$. Then, $\bm{\Psi}_{\mathcal{P}}^T$ is a normalized version of $\bm{Z}_{\mathcal{P}}$ and the columns of $\bm{\Psi}_{\mathcal{P}}\in\mathbb{R}^{n_r\times n_t}$ describe how each of the coefficients in a reduced representation $\bm{z}\in\mathbb{R}^{n_r}$ change to describe all the snapshots. We now have: 

\begin{equation}
	\label{SVD}
	\bm{Z}_{\mathcal{P}}=\bm{{\Phi}_{\mathcal{P}}^T \bm{D}=\bm{\Sigma}_{\mathcal{P}} \bm{\Psi}_{\mathcal{P}}^T} \rightarrow \tilde{\bm{D}}=\bm{\Phi}_{\mathcal{P}}\bm{\Sigma}_{\mathcal{P}}\bm{\Psi}_{\mathcal{P}}^T\,.
\end{equation}

Equation \eqref{SVD} is a truncated singular value decomposition (SVD) of the snapshot matrix. Combining this with \eqref{EIG_POD}, one can see that $\bm{\Sigma}_{\mathcal{P}}=\bm{\Lambda}_{\mathcal{P}}^{1/2}$ and that the columns of $\bm{\Psi}$ must also be orhonormal, i.e. $\bm{\Psi}_{\mathcal{P}}^T\bm{\Psi}_{\mathcal{P}}=\bm{I}$. These are eigenvectors of another important matrix, i.e. 

\begin{equation}
	\label{EIG_sPOD}
	\bigl(\bm{D}^T \bm{D} \bigr) \bm{\Psi}_{\mathcal{P}}= \bm{\Psi}_{\mathcal{P}} \bm{\Lambda}_{\mathcal{P}}\,,
\end{equation} with eigenvalues shared with \eqref{EIG_POD}. In what follow, we define $\bm{K}:=\bm{D}^T\bm{D}$. In fluid dynamics, $\bm{K}$ is the temporal correlation matrix and computing 
the reduced representation $\bm{Z}$ from $\bm{\Psi}$ using \eqref{EIG_sPOD} (thus avoiding computing $\bm{\Phi}$) is known as `snapshot POD' \citep{Holmes1996}. In machine learning, $\bm{K}$ is the Gram matrix and the snapshot POD is known as `dual PCA' \citep{Ghojogh2019PCA}. Moreover, in fluid dynamics, the columns of $\bm{\Phi}$ are \emph{spatial structures} while the columns of $\bm{\Psi}$ are \emph{temporal structures} of each mode.  In machine learning, the columns of $\bm{\Phi}_P$ are \emph{principal components} while the rows of $\bm{Z}_P$ are referred to as \emph{scores}.



The two alternatives of encoding/decoding yields two ways of autoencoding the same dataset, i.e. $\tilde{\bm{D}}=\bm{\Phi}_{\mathcal{P}}\bm{\Phi}_{\mathcal{P}}^T \bm{D}$ or $\tilde{\bm{D}}=\bm{D}\bm{\Psi}_{\mathcal{P}}\bm{\Psi}_{\mathcal{P}}^T$.
Therefore, the fact that this is the linear autoencoder that minimizes \eqref{COST} implies that the matrices $\bm{\Phi}\bm{\Phi}^T$ and $\bm{\Psi}\bm{\Psi}^T$ are as close as possible to the identity matrices. Note now that the temporal correlation matrix in the reduced space is:

\begin{equation}
	\label{TEMP_Z}
	\bm{Z}_{\mathcal{P}}^T\bm{Z}_{\mathcal{P}}=\bigl(\bm{\Phi}_{\mathcal{P}}^T \bm{D}\bigr)^T\bigl(\bm{\Phi}_{\mathcal{P}}^T \bm{D}\bigr)=\bm{D}^T \bm{\Phi}_{\mathcal{P}} \bm{\Phi}_{\mathcal{P}}^T \bm{D}\,.
\end{equation} 

Thus we see that the temporal correlation matrix in the reduced space ($\bm{Z}_{\mathcal{P}}^T\bm{Z}_{\mathcal{P}}$) is as close as possible to the one in the original space ($\bm{D}^T \bm{D}$) if $\bm{\Phi}_{\mathcal{P}}\bm{\Phi}_{\mathcal{P}}^T$ is as close as possible to the identity matrix. We have entered manifold learning: the temporal correlation matrix (or Gram matrix) can be seen as a measure of similarity, i.e. $	\mathcal{S}_{d}(\bm{d}_k,\bm{d}_j)=\bm{d}^T_j\bm{d}_k$ and $ \mathcal{S}_z(\bm{z}_k,\bm{z}_j)=\bm{z}^T_j\bm{z}_k $, and we have just shown that the reduced representations in the PCA/POD is such that $\mathcal{S}_{d}\approx \mathcal{S}_{z}$ as much as possible.

We can now move to the final step. Let $r_{i,j}$ denote the squared Euclidean distance between two snapshots $\bm{d}_{i}$ and $\bm{d}_j$. We thus have: 

\begin{equation}
	\label{Dist_EQ}
	r_{i,j}=||\bm{d}_i-\bm{d}_j||^2_2=\bm{d}_i^T\bm{d}_i-2\bm{d}^T_i\bm{d}_j+\bm{d}^T_j\bm{d}_j\,.
\end{equation} 

Denoting as $\bm{R}$ the matrix collecting all the square distances, i.e. $\bm{R}[i,j]=r_{i,j}$, and defining $\bm{k}:=\mbox{diag}(\bm{K})$, equation \eqref{Dist_EQ} can be written as $\bm{R}=\bm{1}\bm{k}-2\bm{K} +\bm{k} \bm{1}^T$, where $\bm{1}\in\mathbb{R}^{n_t}$ is the vector of ones. Because the matrix of distances $\mathbf{R}$ is invariant to shifts of the origin, it is worth centering the data by removing the average rows and the average columns of $\mathbf{R}$. This operation is called double-centering and can be performed using the centering matrix $\bm{H}:=\bm{I}-1/n_t\bm{1}\bm{1}^T\in\mathbb{R}^{n_t}$, so that the double centered distance matrix is $\bm{H}\bm{R}\bm{H}$. It is now possible to derive the following important relation between $\bm{K}$ and $\bm{R}$ (see \cite{Ghojogh2020}):

\begin{equation}
	\label{FUND}
	\bm{H}\bm{K}\bm{H} = -\frac{1}{2}\bm{H}\bm{R}\bm{H}\,.
\end{equation}

Noticing that centering operation is idempotent, if the data matrix has been centered (that is the temporal average has been removed to all columns), then $\bm{D}\leftarrow \bm{D}\bm{H}$, and $\bm{K}\leftarrow \bm{H}\bm{K}\bm{H}$. We thus have a connection between a measure of similarity in terms of \emph{correlation} or \emph{angles} (i.e. $\bm{K}$) and a measure of similarity in terms of \emph{distances} (i.e. $\bm{R})$.

For a centered dataset, if the Euclidean inner product is used to measure angles and the Eulerian distance is used to measure distances, preserving angles is equivalent to preserving distances. Computing $\mathbf{K}$ from \eqref{FUND} and proceeding with the snapshot POD leads to the `classic' Multidimensional Scaling (MDS) by \cite{Torgerson1952}. But the idea can be generalized. The Kernel PCA introduces a different way of measuring angles while the ISOMAP introduces a different way of measuring distances. 
To conclude this section, the following Python function can be used to compute the $n_r$ dimensional POD encoding $\bm{Z}_{\mathcal{P}}$ for the data matrix $\mathbf{D}$ using the snapshot-based approach:

	\begin{centering}
	\begin{lstlisting}[language=Python,linewidth=15.5cm,xleftmargin=.05\textwidth,xrightmargin=.05\textwidth,backgroundcolor=\color{yellow!10}]
def PCA_TRANS(D,n_r=2):
 # Compute Temporal Correlation Matrix  K
 K=D.T@D; n_t=np.shape(D)[1]
 # Diagonalize and Sort
 lambdas,Psi_P=linalg.eigh(K,subset_by_index=[n_t-n_r, n_t-1]) 
 lambdas,Psi_P=lambdas[::-1],Psi_P[:,::-1]  
 # Take the first n_r modes
 Sigma_P=np.sqrt(lambdas[:n_r]); Psi_P=Psi_P[:,:n_r] 
 # Encode
 Z_P=np.diag(Sigma_P)@Psi_P.T 
 return Z_P, Sigma_P
	\end{lstlisting}
\end{centering}

In the previous code, it is assumed that the package package \textit{linalg} is importent from scipy \citep{2020SciPy-NMeth} and numpy \citep{harris2020array} is imported as np. A computationally more efficient approach consist in using the randomized SVD \citep{Halko2011}, available in the scikit-learn \citep{scikit-learn}:

	\begin{centering}
	\begin{lstlisting}[language=Python,linewidth=15.5cm,xleftmargin=.05\textwidth,xrightmargin=.05\textwidth,backgroundcolor=\color{yellow!10}]
def PCA_TRANS_rand(D,n_r=2):
  # Compute Temporal Correlation Matrix  K
  K=D.T@D
  #Diagonalize using truncated SVD
  svd = TruncatedSVD(n_r); svd.fit_transform(K)
  Psi_P = svd.components_.T; Lambda_P=svd.singular_values_
  Sigma_P=np.sqrt(Lambda_P)
  # Encode
  Z_P=np.diag(Sigma_P)@Psi_P.T 
  return Z_P, Sigma_P
	\end{lstlisting}
\end{centering}

This approach exploits the fact that the SVD of $\mathbf{K}$ is equivalent to its eigenvalue decomposition and the randomized approach is much faster. However, it is also numerically less accurate. The randomised approach can be used to replace the eigenvalue decomposition of hermitian matrices also in the kPCA and ISOMAPs.

\subsection{From POD (PCA) to kernel PCA (POD) }\label{sec:3p2}

The kernel (kPCA) \citep{Schoelkopf1997} is a \emph{kernelized} version of the PCA which modifies the left hand side of \eqref{FUND} by replacing the temporal correlation (Gram) matrix with a kernel matrix. The underlying idea is to perform the PCA on a dataset which has been first transformed by a nonlinear function $\xi:\mathbb{R}^{n_s}\rightarrow\mathbb{R}^{n_f}$. This function brings the data onto a higher (potentially infinite) dimensional space called \emph{feature} space.

The motivation for this transformation is that linear operations in the feature space (e.g. projections) are nonlinear in the original space, and the function $\xi$ can (hopefully) highligh special features in the data. The PCA in this nonlinear space is the kernel PCA.

Let us briefly review the idea behind the kernelization, which is performed using the popular \emph{kernel trick} \citep{Schoelkopf1997}. This consists in avoiding operations in the feature space (and even an explicit definition of $\xi$) by using a function, called \emph{kernel} function $\kappa(\bm{d}_i,\bm{d}_j):\mathbb{R}^{n_f}\times\mathbb{R}^{n_f}\rightarrow \mathbb{R}$, to compute the inner products in $\mathbb{R}^{n_s}$. Thus we have:

\begin{equation}
\kappa(\bm{d}_i,\bm{d}_j)=\langle \xi(\bm{d}_i),\xi(\bm{d}_j) \rangle =\xi^T(\bm{d}_i)\xi(\bm{d}_j)\,.
\end{equation}

In this work, we consider a Gaussian kernel $\kappa(\bm{d}_i,\bm{d}_j)=\exp(-\gamma r_{i,j})$, with $\gamma>0$ an hyperparameter. This is usually of the order $\gamma \sim 1/n_s$ or can be set by fixing the least value of $\kappa(\bm{d}_i,\bm{d}_j)$. Denoting as $r_M=\max(r_{i,j})$ the largest squared Euclidean distance in the dataset and as $\kappa_m$ the lowest limit for the kernel function, one has $\gamma=-\ln(\kappa_m)/r_M$.

Defining $\bm{K}_{\xi}:=\xi(\bm{D})^T\xi(\bm{D}) \in\mathbb{R}^{n_t\times n_t}$, one thus has $\bm{K}_{\xi}[i,j]=\kappa(\bm{d}_i,\bm{d}_j)$. Furthermore, let $\bm{C}_{\xi}:=\xi(\bm{D})\xi(\bm{D})^T \in\mathbb{R}^{n_f\times n_f}$ and let $\bm{\Phi}_{\xi}=[\phi_{\xi1},\dots\phi_{\xi n_f}]$ be the set of principal components in the feature space. Approximating these as linear combinations of the features gives

\begin{equation}
	\label{eq8}
	\phi_{\xi r}=\sum^{n_p}_{i=1}\bm{\Psi}_{\xi}[{r,i}]\xi(\bm{d}_i)\rightarrow \bm{\Phi}_\xi=\xi(\bm{D})\,\bm{\Psi}_{\xi}\,.
\end{equation} 

Then, multiplying the eigenvalue problem for $\bm{K}_{\xi}$ by $\xi^T(\bm{D})$ and introducing \eqref{eq8} gives

\begin{equation}
	\label{eq9}
		\xi(\bm{D})^T(\xi(\bm{D})\xi(\bm{D})^T) \xi(\bm{D})\bm{\Psi}_{\xi} = \xi(\bm{D})^T \xi(\bm{D}) \bm{\Psi}_{\xi} \bm{\Lambda}_{\xi}\rightarrow \bm{K}_{\xi}^2\bm{\Psi}_{\xi}
		=\bm{K}_{\xi}\bm{\Psi}_{\xi}\bm{\Lambda}_{\xi}.
\end{equation}

As long as $\bm{K}_{\xi}$ is non-singular, this results in an eigenvalue problem for $\bm{K}_{\xi}$. Therefore, the kernelized version of equation \eqref{SVD}, that is the kPCA encoder, is

\begin{equation}
	\label{SVD_xi}
\bm{Z}_{\xi}=\bm{{\Phi}_{\xi}^T \xi(\bm{D})=\bm{\Sigma}_{\xi} \bm{\Psi}_{\xi}^T} \quad \mbox{or, equivalently,}\quad \bm{Z}_{\xi}=\bm{\Psi}_{\xi} \,\bm{K}_{\xi}\,,
\end{equation} with $\bm{\Sigma}_{\xi}=\bm{\Lambda}^{1/2}_{\xi}$.

We thus see that the encoding algorithm is identical to the snapshot POD once the temporal correlation matrix is replaced by the kernel matrix. On the other hand, the decoding is a much more complex (and generally not well posed) problem. The `natural' decoder requires identifying both $\xi()$ and the eigenfunctions $\Phi_{\xi}$, to construct the truncated SVD in the feature space, and the inverse $\xi^{-1}()$ to map the approximation from the feature space to the original space. None of these operation is usually possible. We return to this point in Section \ref{sec:3p5}.


The python function below can be used to compute the $n_r$ dimensional encoding $\bm{Z}_{\xi}$ using a Gaussian kernel with $\gamma$ given by the minimal kernel value $\kappa_m$

\begin{centering}
	\begin{lstlisting}[language=Python,linewidth=16.5cm,xleftmargin=.05\textwidth,xrightmargin=.05\textwidth,backgroundcolor=\color{yellow!10}]
def kPCA_TRANS(D,n_r=2,k_m=0.1,cent='y'):
 # Compute Eucledean distances
 r_ij=pdist(D.T,'sqeuclidean'); R=squareform(r_ij) 
 # Look for the largest distance and define gamma:
 M=np.max(R); gamma=-np.log(k_m)/M    
 # Compute the Kernel Matrix
 K_zeta=np.exp(-gamma*R) ; n_t=np.shape(D)[1]
 # Center the Kernel Matrix (if cent is 'y'):
 if cent =='y':
  H=np.eye(n_t)-1/n_t*np.ones_like(K_zeta)   
  K_zeta=H@K_zeta@H.T    
 # Diagonalize and Sort
 lambdas,Psi_xi=linalg.eigh(K_zeta,subset_by_index=[n_t-n_r,n_t-1]) 
 lambdas,Psi_xi=lambdas[::-1],Psi_xi[:,::-1]; Sigma_xi=np.sqrt(lambdas); 
 # Encode
 Z_xi=np.diag(Sigma_xi)@Psi_xi.T 
 return Z_xi, Sigma_xi
	\end{lstlisting}
\end{centering}

The functions \textit{pdist} and \textit{squareform} are imported from \textit{scipy.spatial.distance}. The last steps in lines 13 to 18 are identical to the POD implementation while the optional argument \textit{cent} allows the user to decide if performing the kernel centering using the matrix $\mathbf{H}$ introduced in Section \ref{sec:3p1}. Such centering might not be relevant in non-stationary datasets.

\subsection{From MDS to ISOMAPs }\label{sec:3p3}

\begin{figure}[ht]
	\centering
		\includegraphics[width=.3\textwidth ]{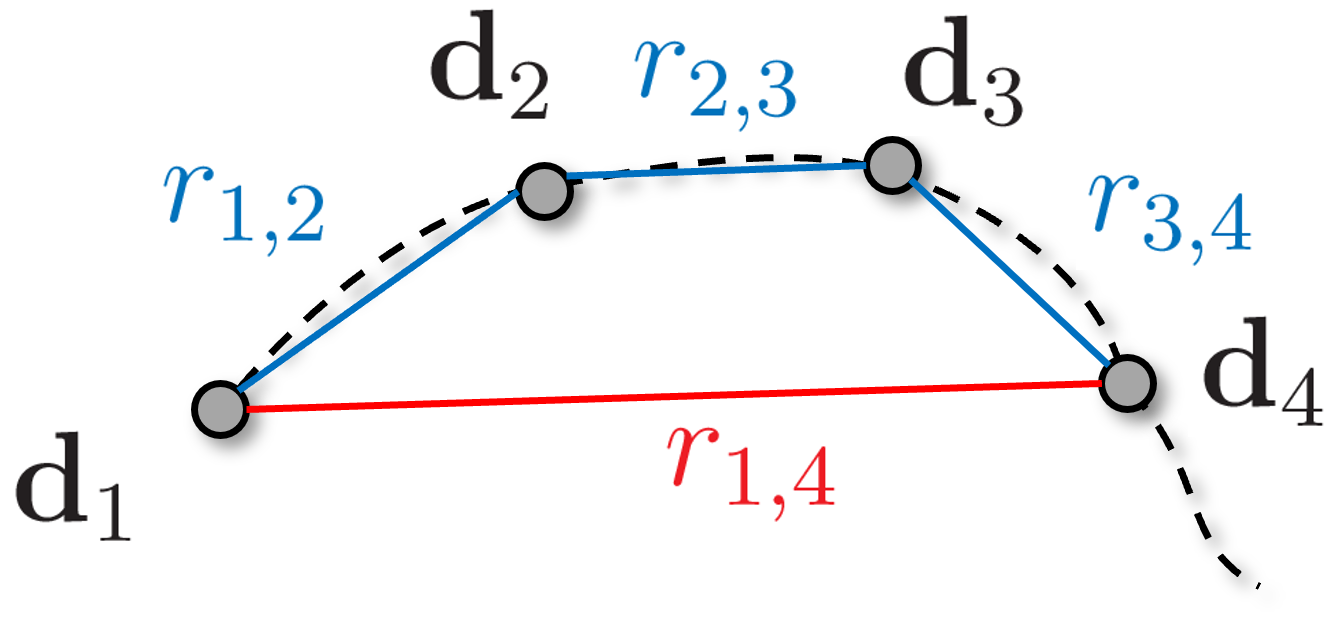}
	\caption{Schematic illustration of the geodesic distance versus Euclidean distance. The Eulcidean distance between the point $\bm{d}_1$ and $\bm{d}_4$ is $r_{1,4}$, shown in red. The (approximated) geodesic distance is $g_{1,4}=r_{1,2}+r_{2,3}+r_{3,4}$.} 
	\label{fig1}
\end{figure}

The isometric feature mapping (ISOMAP, \cite{Tenenbaum2000}) modifies the right hand side of \eqref{FUND} by replacing the Euclidean distance with an approximation of the geodesic distance. This is the shortest path between two points that preserve the topology of the manifold over which they lay. This notion is illustrated in the sketch in Figure \ref{fig1}.

Let $\bm{d}_1,\bm{d}_2,\bm{d}_3 \bm{d}_4$ be four points in $\mathbb{R}^2$, laying on a curved manifold (shown with dashed lines). The Euclidean distance $r_{1,4}$ ignores the shape of the manifold while the geodesic distance is computed \emph{on} the manifold. The approximated geodesic distance in this example is $g_{1,4}=r_{1,2}+r_{2,3}+r_{3,4}$.

The approximated geodesic distance is also referred to as \emph{curvilinear} distance \citep{LEE} and can be approximated by summing the next-neighbours distances. The identification of the neighbors for each snapshot can be carried out using classic k-neirest neibhours algorithms \citep{Rivest2009}. Assuming, for the sake of simplicity, that the data has been centered, on can identify the kernel matrix $\bm{K}_{I}$:

\begin{equation}
	\label{K_I}
	\bm{K}_I=-\frac{1}{2}\bm{H} \bm{G} \bm{H}\,,
\end{equation} with $\bm{G}[i,j]=g_{i,j}$ the matrix collecting the approximated geodesic distances. The reader is referred to \cite{Choi2004} for the kernel formulation of the ISOMAP, to \cite{Ghojogh2021C} for an analysis of the links between kPCA and ISOMAPs and to \cite{Cox2008} for a discussion on the centering of this kernel matrix. Once $\bm{K}_I$ is computed, the ISOMAP encoding proceeds like the snapshot POD, that is 

\begin{equation}
	\label{ISO_ENC}
\bm{Z}_{I}=\bm{\Sigma}_{I} \bm{\Psi}_{I}^T \quad \mbox{with}\quad \bm{K}_{I}\bm{\Psi}_{I}=\bm{\Sigma}_I \bm{\Psi}_{I} \,.
\end{equation}

Concerning the decoding, like the kPCA and all manifold learning techniques, the problem is ill posed. We return to this point in section \ref{sec:3p5}. The script that follows provides a python function to compute the $n_r$ dimensional encoding $\bm{Z}_{I}$ for an ISOMAP with $k$ nearest neighbours

\begin{centering}
	\begin{lstlisting}[language=Python,linewidth=16.5cm,xleftmargin=.05\textwidth,xrightmargin=.05\textwidth,backgroundcolor=\color{yellow!10}]
def ISO_TRANS(D,n_r=2,k=5):
 # Compute k neirest neibhbors distances
 nbrs_ = NearestNeighbors(n_neighbors=k); nbrs_.fit(D.T)
 # Compute the neighbors graph
 nbg = kneighbors_graph(nbrs_,n_neighbors=k,p=2,mode="distance")
 # Commpute the shortest path on the graph
 R_G= shortest_path(nbg,directed=False)
 # Get the matrix G:
 G=-0.5*R_G**2    
 # Center the matrix:
 H=np.eye(n_t)-1/n_t*np.ones_like(G)   
 K_I=H@G@H.T    
 # Diagonalize and Sort
 lambdas,Psi_I=linalg.eigh(K_I,subset_by_index=[n_t-n_r, n_t-1]) 
 lambdas,Psi_I=lambdas[::-1],Psi_I[:,::-1] ; Sigma_I=np.sqrt(lambdas); 
 # Encode
 Z_I=np.diag(Sigma_I)@Psi_I.T 
 return Z_I, Sigma_I
	\end{lstlisting}
\end{centering}

The functions \textit{NearestNeighbors} and \textit{kneighbors\_graph} are imported from \textit{sklearn.neighbors} in the scikit-learn \citep{scikit-learn} library. The function \textit{shortes\_path} is imported from \textit{scipy.sparse.csgraph}. These functions have many options on the metrics and the algorithms used to construct the graphs; the provided function, which uses default settings, serves purely illustrative purposes. The reader is referred to the documentation for more information and is invited to consult the more sophisticated implementation in scikit-learn.

\subsection{From MDS to LLE}\label{sec:3p4}

The Locally Linear Embedding (LLE, \cite{LLE_Ref}) is conceptually similar to ISOMAPS, but it is a `local' approach: the goal is preserving distances between neighbours and not within the whole set of snapshots. If neighbours are close enough, the geodesic distance and the Euclidean distances are similar (hence ISOMAPS reduces to MDS), and the underlying manifold can be approximated as a set of linear patches. This makes the algorithm more efficient than ISOMAP in that it avoids computing distances between points that are far apart.

The algorithm is composed of three steps. The first step is a k-nearest neighbour search like in ISOMAPS. This produces a matrix $\bm{N}_j \in\mathbb{R}^{n_s\times k}$ collecting the $k$ nearest neighbours for the snapshot $\bm{d}_j$. The second step consists in a locally linear fit: we seek to approximate every snapshot as a linear combination of its neighbours. Therefore, we minimize the following

\begin{equation}
	\label{J_LLE}
	J_1(\mathbf{W})=\sum^{n_t}_{j=1}||\bm{d}_j-\bm{N}_j \bm{w_j}||^2_2\,,
\end{equation} where $\mathbf{W}\in\mathbb{R}^{n_t\times k}=[\bm{w}_1,\cdots \bm{w}_{n_t}]^T$ is the matrix collecting the weights for each snapshot $\bm{w}_j$ (stacked row-wise). This minimization is constrained by the condition $\bm{1}^T\bm{w}_j=1$, that is the sum of weights for each snapshot must be unitary. This minimization admits a closed form solution (see \cite{Ghojogh2020a}), which reads

\begin{equation}
	\label{W_LLE}
	\bm{w}_{j}=\frac{\bm{L}^{-1}_j \bm{1}}{\bm{1}^T\bm{L}^{-1}_{i}\bm{1}} \quad \mbox{with} \quad \bm{L}_i:= (\bm{d}_j\bm{1}-\bm{N}_j)^T (\bm{d}_j\bm{1}-\bm{N}_j)\in\mathbb{R}^{k\times k}\,.
\end{equation}

The matrix $\bm{L}_j$ is the correlation matrix centered at each snapshot $j$ and measuring the correlation between neighbours. Its small size makes the inversion computationally irrelevant.

Finally, the last step consists in finding the reduced representation that preserves the previously derived locally linear approximation. We thus need the compressed representation $\bm{Z}_I\in\mathbb{R}^{n_r\times n_t}$ minimizing the following cost function:

\begin{equation}
	\label{Z_L}
	J_2(\bm{Z}_{L}) =\sum^{n_t}_{j=1}||\bm{Z}_L \bm{\delta}_j -\bm{Z}_{L}  \bm{w_j}||^2_2 = ||\bm{Z}_I-\bm{Z}_I\tilde{\bm{W}}||_F\,,
\end{equation} where $\bm{\delta}_j\in\mathbb{R}^{n_t}$ is the delta vector equal to $1$ at the j-th entry and zero everywhere else, $||\bullet||_F$ is the Frobenious norm and $\tilde{\bm{W}}\in\mathbb{R}^{n_t\times n_t}$ is the agumented matrix of weights, with entries equal to the previously computed weights $\bm{w}_j[i]$ if the snapshot $j$ is among the neibhours of $i$ and zero if it is not.
 Also this minimization admits a closed form solution, which leads to an eigenvalue problem (see \cite{Ghojogh2020a}):

\begin{equation}
	\label{LLE_SOL}
	\bm{M}\bm{Z}^T_L=\bm{Z}^T_L\bm{\Lambda}_L \quad \mbox{with} \quad \bm{M}:= (\bm{I}-\tilde{\bm{W}})^T (\bm{I}-\tilde{\bm{W}})\in\mathbb{R}^{n_t\times n_t}\,.
\end{equation} 

However, differently from the previous method, the minimization in \eqref{Z_L} is provided by the eigenvectors linked to the \emph{smallest} eigenvalues of $\bm{M}$. The following python function can be used to compute the $n_r$ dimensional encoding $\bm{Z}_L$ using the LLE with $k$ neirest neibhours:

\begin{centering}
	\begin{lstlisting}[language=Python,linewidth=16.5cm,xleftmargin=.05\textwidth,xrightmargin=.05\textwidth,backgroundcolor=\color{yellow!10}]
def LLE_TRANS(D,n_r=3,k=5):
 # Compute k neirest neibhbors distances
 nbrs_ = NearestNeighbors(n_neighbors=k+1); nbrs_.fit(D.T)
 # Compute the Matrix W (and make it dense)
 W = barycenter_kneighbors_graph(nbrs_,n_neighbors=k); n_t=np.shape(D)[1]
 # Compute Matrix M= (I-W)^T(I-W) in sparse format 
 M=(eye(n_t)-W).T@(eye(n_t)-W)
 # Diagonalize and Sort (use sparse eig)
 lambdas,Psi_L=eigsh(M, n_r + 1, sigma=0.0)
 # Remove the lowest eigenvalue 
 return Psi_L[:,1:].T,lambdas[1:]
	\end{lstlisting}
\end{centering}
 
This function uses \textit{barycenter\_kneighbors\_graph
} from sklearn.manifold's \textit{\_locally\_linear} and the functions \textit{eye} and \textit{eigsh} from scipy's \textit{sparse.linalg}. The first is used to compute the matrix $\bm{W}$ using an efficient implementation of \eqref{W_LLE}. This is returned as a sparse array. \textit{eye} creates an identity matrix as a sparse object while \textit{eigsh} is a sparse eigenvalue solver for hermitian matrices. This solver allows to look for eigenvalues near a user defined value \textit{sigma} (in this case 0). The first eigenvalue is usually zero to machine precision and associated to a constant eigenvector; both are removed in the return line.

\subsection{A note on the Decoding Process}\label{sec:3p5}

Finding the decoding function in nonlinear manifold learning is an ill posed problem. In the vast literature on kPCA, this is known as the `pre-image' problem (see \cite{MIKA,Bakir2003,Bakir2004,Bakir2007,Kwok2004,Zheng2006,Honeine2011}. The most common solution consists in using an interpolation or a regression approach. Given a set of snapshots $\bm{d}_j\in\mathbb{R}^{n_s}$ and their associated reduced representation $\bm{z}_j\in\mathbb{R}^{n_r}$, finding a function such that $\bm{d}_j\approx\mathcal{D}(\bm{z}_j)$ is a supervised learning problem for which machine learning offers an arsenal of tools. In kernel PCA, the most common is Kernel Ridge regression/interpolation \citep{Bakir2003,Bakir2004} using the same kernel function used in the encoding. More sophisticated methods are reviewed by \cite{Honeine2011}.

In LLE, a natural approach is to use linear regression with the same weights computed during enconding process, as proposed by \cite{Lawrence_SAUL}. Given a vector $\bm{z}_l\in\mathbb{R}^{n_r}$, using a k-nearest neighbor algorithm to derive the set of weights such that $\bm{z}_l\approx \bm{P}_l \bm{w}_l$, with $\bm{P}_l\in \mathbb{R}^{n_r\times k}$ the matrix collecting the k-neirest neighbors of $\bm{z}_l\in\mathbb{R}^{n_r}$, the associated high dimensional counterpart (pre-image) is $\tilde{\bm{d}}_l=\bm{N}_l \bm{w}_l$, where $\bm{N}_l\in \mathbb{R}^{n_s\times k}$ is the matrix collecting the k neirest neighbors of $\tilde{\bm{d}}_l\in\mathbb{R}^{n_s}$. This approach was also used with excellent results by \cite{Ehlert2019}. An elegant variant, based on a first order Taylor expansion of the decoder function, is presented by \cite{Farzamnik2022}.

In this work, aiming at a first overview of these methods, we rely on k-neirest neighbors linear regression because it is the simplest (thus also less prone to over-fitting) and computationally cheaper. Given set of weights in \eqref{W_LLE} and given $\tilde{\bm{W}}\in\mathbb{R}^{n_t\times n_t}$ the associated augmented matrix of weights, the decoder is simply $\tilde{\bm{D}}=\bm{D}\tilde{\bm{W}}$. Interpolation is prevented by construction, since the diagonal entries of $\tilde{\bm{W}}$ are zero.

The following script provide a function to compute the decoding as a linear combination of the k neirest neighbors of each snapshot, identified from the reduced dimensional representation:

\begin{centering}
	\begin{lstlisting}[language=Python,linewidth=16.5cm,xleftmargin=.05\textwidth,xrightmargin=.05\textwidth,backgroundcolor=\color{yellow!10}]
def Decode(Z,D,k=10,reg=1e-6):
 # compute the W matrix   
 W=barycenter_kneighbors_graph(Z,n_neighbors=k,reg=reg)
 # Compute the approximation   
 D_tilde =(W @ D.T).T
 return D_tilde
	\end{lstlisting}
\end{centering}	

As a note of warning, it is important to note the performance of any approximated decoding methods based on interpolation should be evaluated on data that has not been used for the encoding process. It is in fact intuitive that a compressed representation $\bm{z}_l$ that is \emph{close} to a vector $\bm{z}_i$ computed by the encoding $\bm{z}_i=\mathcal{E}(\bm{d}_i)$ has a pre-image that is \emph{close} to $\bm{d}_i$ regardless of how well the encoding was performed. This might lead to the misleading conclusion that an autoencoder is lossless if the decoding is carried out on the same data that was encoded. To experiment with this potential problem, the reader is invited to perform kPCA and its inverse using scikit-learn and monitor the reconstruction error on the training data for different values of $n_r$.

Finally, we measure the performances of the autoencoding using two metrics, defined as 
\begin{equation}
	\label{ERR_l2}
	E=\frac{||\bm{D}-\tilde{\bm{D}}||_2}{||\bm{D}||_2}\quad \mbox{and}\quad R_V=1-\rho^2(\bm{G},\bm{R}_Z)\,.
\end{equation}

The first is the normalized $l_2$ error, which the POD/PCA seeks to minimize. The second is an adaptation of the residual variance by \cite{Tenenbaum2000}, as proposed by \cite{Farzamnik2022}. Here $\bm{R}_Z$ denotes the matrix of Euclidean distances in the reduced domain while $\rho(\bm{A},\bm{B})$ denotes the Pearson's correlation coefficient between two matrices $\bm{A}$,$\bm{B}$ (computed entry by entry). If $\rho\approx 0$, the Euclidean distances in the reduced space are well correlated with the geodesic distances in the original space. One could thus expect a better mapping of the underlying manifold.

\section{Selected Test Cases}\label{sec:4}
    
\subsection{Test Case 1: Background Removal in PIV Images}\label{sec4p1}

We consider a filtering problem as a first test case. We test the nonlinear dimensionality reduction techniques described in the previous section on the background removal in PIV images. This problem is extensively described in \cite{MENDEZ2017181}, where the implementation of a POD (PCA)-based framework was first proposed for this problem. This is a special kind of denoising problem and nonlinear methods such as kPCA have been widely used for similar tasks \citep{MIKA,Honeine2011}.

In the PIV background removal, the dataset at hand is a video sequence $\bm{D}\in\mathbb{R}^{n_s\times n_t}$ composed of $n_t$ gray scale images of resolution $n_s$ (every image is reshaped into a column of $\bm{D}$). 
We assume that this dataset is the sum of an `ideal sequence', here denoted as $\bm{D}_P$, containing the scattered light from particles, and a `background sequence', here denoted as $\bm{D}_B$, containing the time-varying sources of background noise (e.g. laser reflections).

 \begin{figure}[ht]
 	\centering
 	\includegraphics[width=.3\textwidth ]{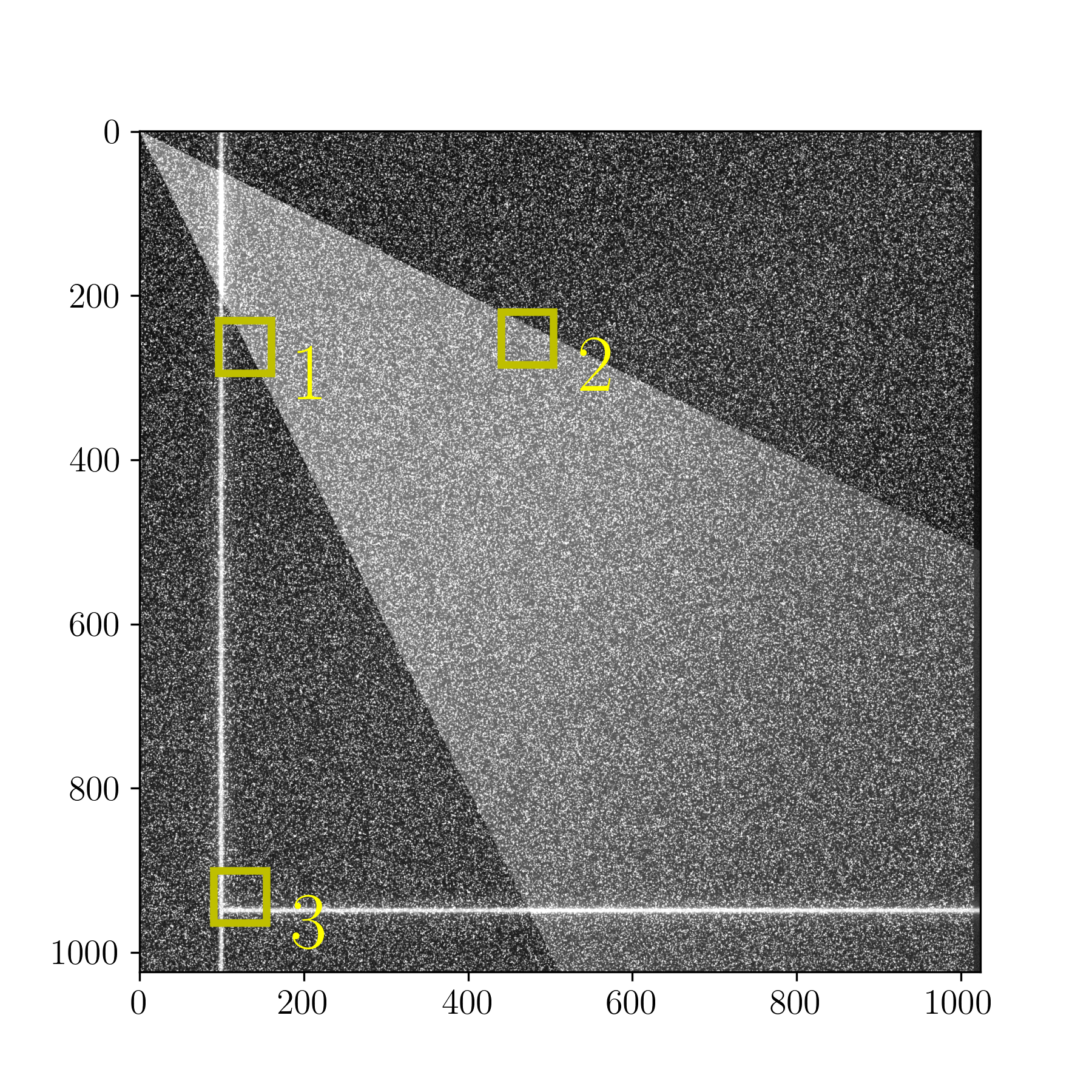}\\
 	
 	\includegraphics[width=0.9\textwidth ]{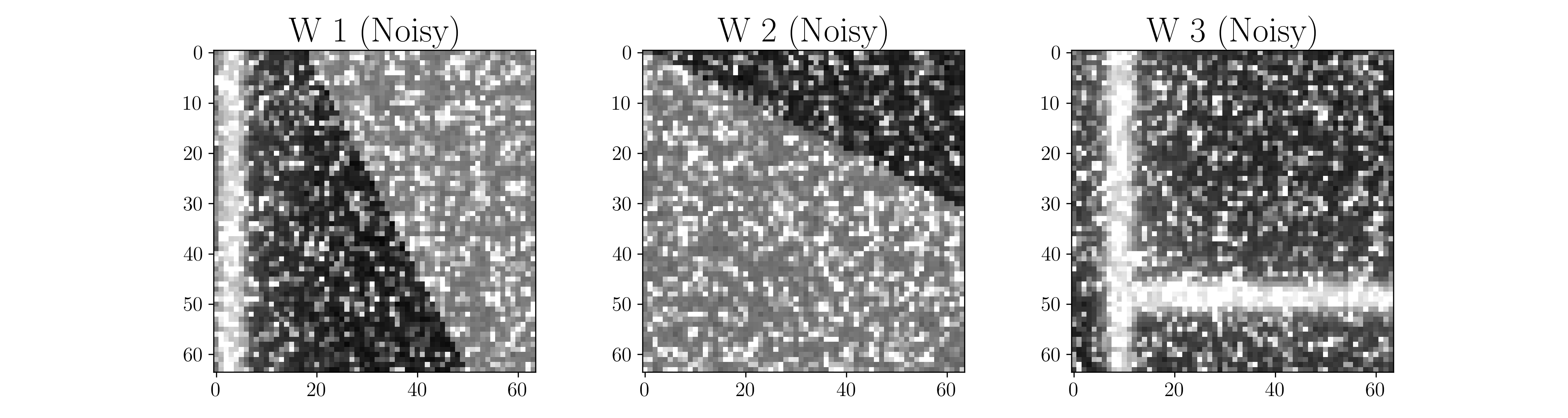}
 	
 	\includegraphics[width=0.9\textwidth ]{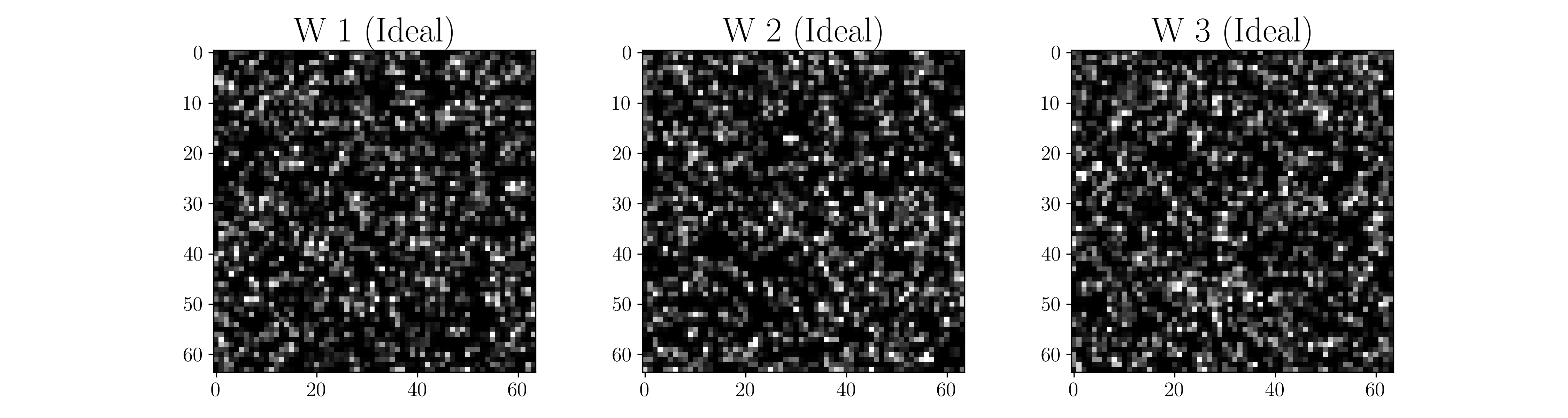}
 	 	
 	\caption{Illustration of the synthetic test case from \cite{MENDEZ2017181} used to test the performances of nonlinear autoencoders in solving the background problem in PIV. The figure on the top shows a snapshot of the video sequence. The figures below show a zoom in three windows of 64x64 pixels (indicated in yellow in the top figures) for the noisy image and the underlying ideal image.} 
 	\label{PIV_IMAGES_TEST_ZOOMS}
 \end{figure}

In the autoencoder formulation of the denoising problem, we hope that the approximated field is $\tilde{\bm{D}}\approx \bm{D}_B$, such that one can filter the video sequence with a simple subtraction $\tilde{\bm{D}}_P=\bm{D}-\tilde{\bm{D}}$. The rationale behind this approach is that the background noise in PIV images tends to be more correlated than PIV particles (which can be well approximated by a sequence of random patterns). Hence, an autoencoder might well approximate the background's evolution and filter less correlated contributions. The idea is common in computer vision, where it is used to distinguish moving objects from static backgrounds \citep{Bouwmans}.
     
The test case selected in this work is the first synthetic case presented in \cite{MENDEZ2017181}, to which the reader is referred for more details. The set of images has been made available at \url{https://osf.io/g7asz/download}. Briefly, this consist of a set of four sources of background noise. A snapshot of this dataset is shown in Figure \ref{PIV_IMAGES_TEST_ZOOMS} on the top. The first two noise sources are triangular areas of non-uniform and time varying illumination, extending from the top corners to the bottom of the image. The third consists of vertical and horizontal thick lines featuring light reflection and flare. The last one consists of Gaussian distributed background noise. The sequence consists of $n_t=100$ images with $n_s=$1024 $\times$ 1024 pixels. The images have an 8-bit dynamic range and are rescaled in the range $[0,1]$. Alghough two of the noise sources have a time-resolved evolution, a random shuffle is applied to simulate a non-time resolved acquisition.  

Three 64 x 64 windows are used to evaluate the filtering performances. These are marked in yellow in the snapshot in Figure \ref{PIV_IMAGES_TEST_ZOOMS}. A zoomed view in these windows is shown in Figure \ref{PIV_IMAGES_TEST_ZOOMS} for both the noisy and the underlying ideal images. The first windows (W1), is influenced by both the ``vertical reflection'' and the light non-uniformity. Since the first reaches saturations in various snapshots, its removal is particularly challenging. This is more severe in the third window (W3) while the background removal is much easier for window (W2) which is far from saturation. 

In what follows, we evaluate the performances of the autoencoders by measuring how much the filtered image resembles the ideal one. For each window, we define and error $E_B$ as the root mean square difference between the filtered image and the ideal one, that is 

\begin{equation}
	\label{E_B}
	E_B=\frac{1}{\sqrt{n_t n_s}}||\bm{D}_p-\tilde{\bm{D}}||_2\,.
\end{equation}

     \subsection{Test Case 2: Turbulent flow past a cylinder in transient conditions}
     \label{sec4p2}
     
We consider the classic flow past a cylinder as a second test case.
This is arguably the most popular test case in the community of reduced order modeling of fluid flows because it is characterized by an intrinsically low dimensionality. At Reynolds numbers $Re=U_{\infty} d/\nu>47$, with $U_{\infty}$ the free stream velocity, $d$ the cylinder diameter and $\nu$ the kinematic viscosity, this configuration produces the well-known von Karman vortex street. This condition is characterized by a regular shedding of vortices and can be well approximated by three POD modes \citep{NOACK2003}. This configuration was also used by \cite{Ehlert2019} to showcase the LLE's capability to identify the underlying manifold in a transient condition from $Re=40$ to $Re=120$.

\begin{figure}[ht]
     	\centering  
     	\begin{subfigure}[\label{Snap_Cyl}]{
     			\includegraphics[width=.55\textwidth ]{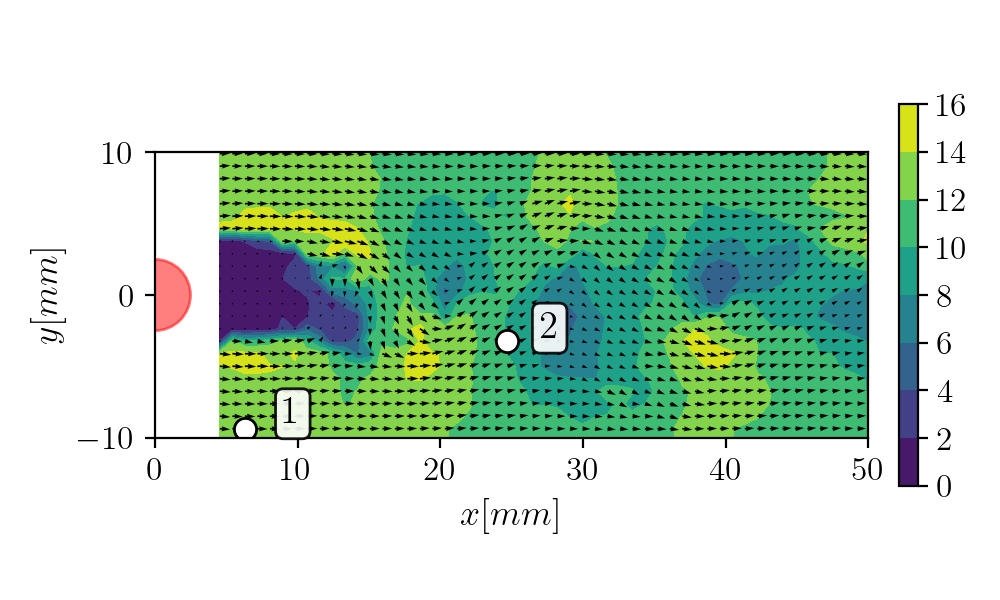}}	
     	\end{subfigure}	    	\\
     	\begin{subfigure}[\label{Plot_1}]{
     			\includegraphics[width=.35\textwidth ]{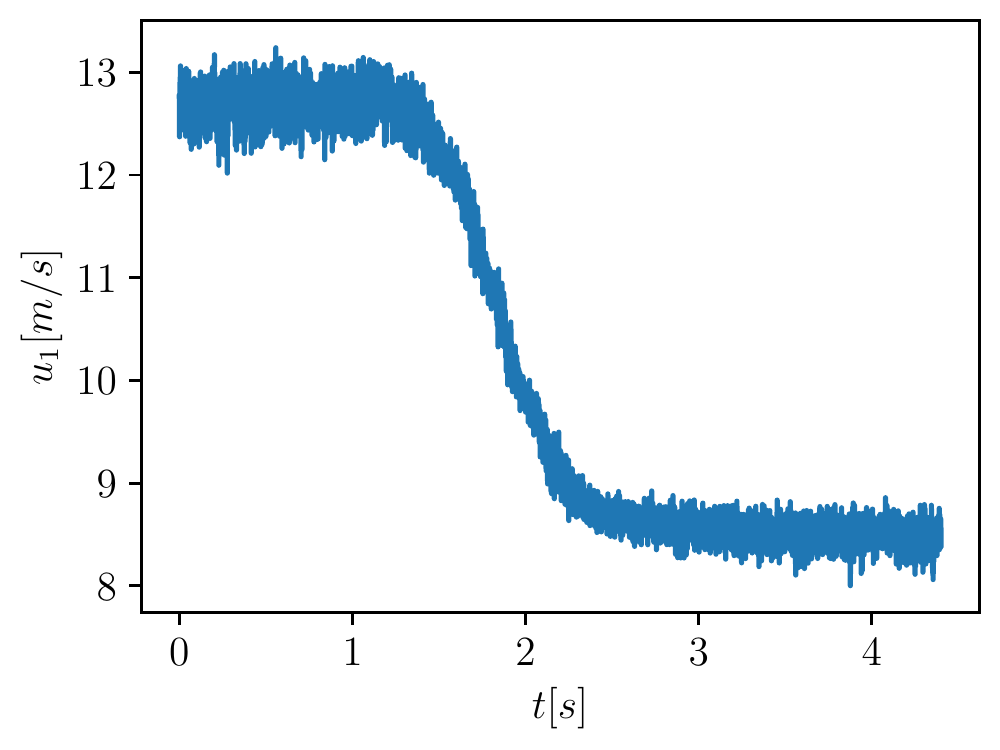}}
     	\end{subfigure}	
     	\begin{subfigure}[\label{Plot_2}]{
     			\includegraphics[width=.35\textwidth ]{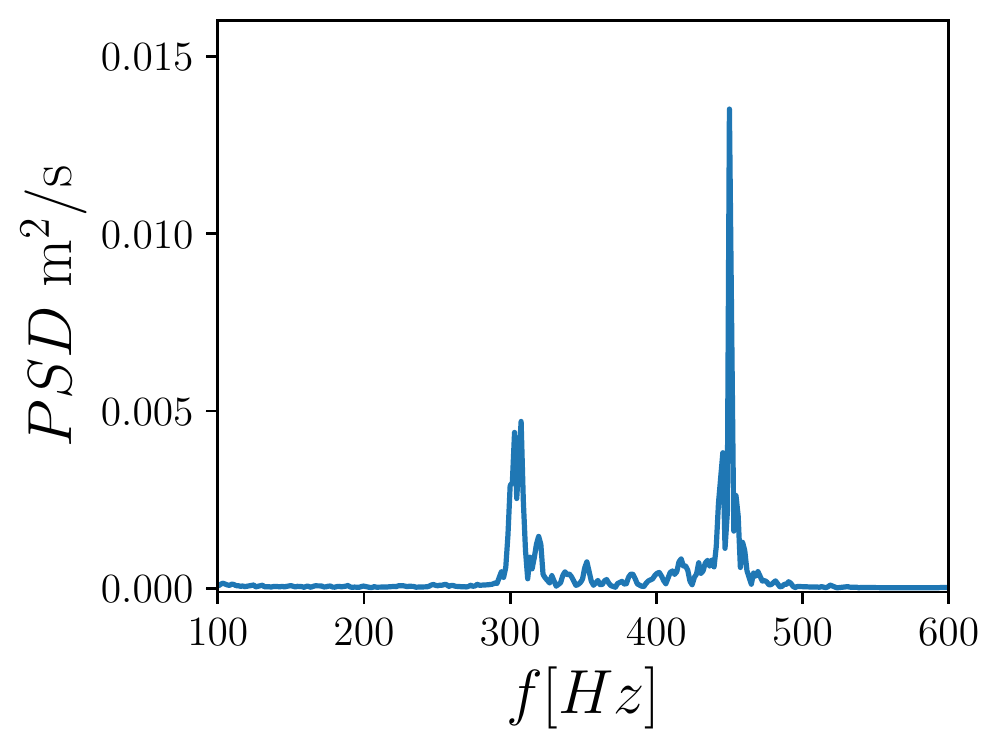}}
     	\end{subfigure}	
     	\caption{a) Snapshot of the velocity field for the second test case: the TR-PIV measurements of the flow past a cylinder. b) Evolution of the free stream velocity (in probe 1), c) power spectral density of the velocity magnitude in probe 2.} 
     	\label{CYL_TEST_CASE}
     \end{figure}

In this test case, we consider a transient condition from $Re\approx4000$ to $Re\approx2600$, using an experimental dataset acquired via Time-Resolved PIV. The experimental set-up and related processing is presented in \cite{Mendez2020}. The dataset consists of $n_t=13200$ velocity fields with $71\times 30$ vectors sampled at $f_s=3$ kHz and has been made available at   \url{https://osf.io/47ftd/download}. A snapshot of the velocity field is shown in Figure \ref{Snap_Cyl}, together with the location of two probes (1) and (2). Figure \ref{Plot_1} shows the evolution of the free stream velocity sampled at probe 1, while Figure \ref{Plot_2} shows the power spectral density of the velocity magnitude in probe (2). Interested readers are referred to \cite{Mendez_Course} for a tutorial on data processing on this test case using Python.

     As visible from the probe 1, the free stream velocity was varied from $U_{\infty}=12$m/s to $U_{\infty}=8$ m/s. Because this variation is sufficiently slow, the flow remains in quasi-steady conditions and the vortex shedding occurs at a constant Strhoual number $St=f d/U_{\infty}\approx 0.19$. Accordingy, the shedding frequency changes from $f\approx 450 $Hz to $f\approx300$ Hz as shown in Figure \ref{Plot_2}.
         


\begin{figure}[ht]
	\centering  
	\begin{subfigure}[\label{Jet_SNAP}]{
			\includegraphics[width=.6\textwidth ]{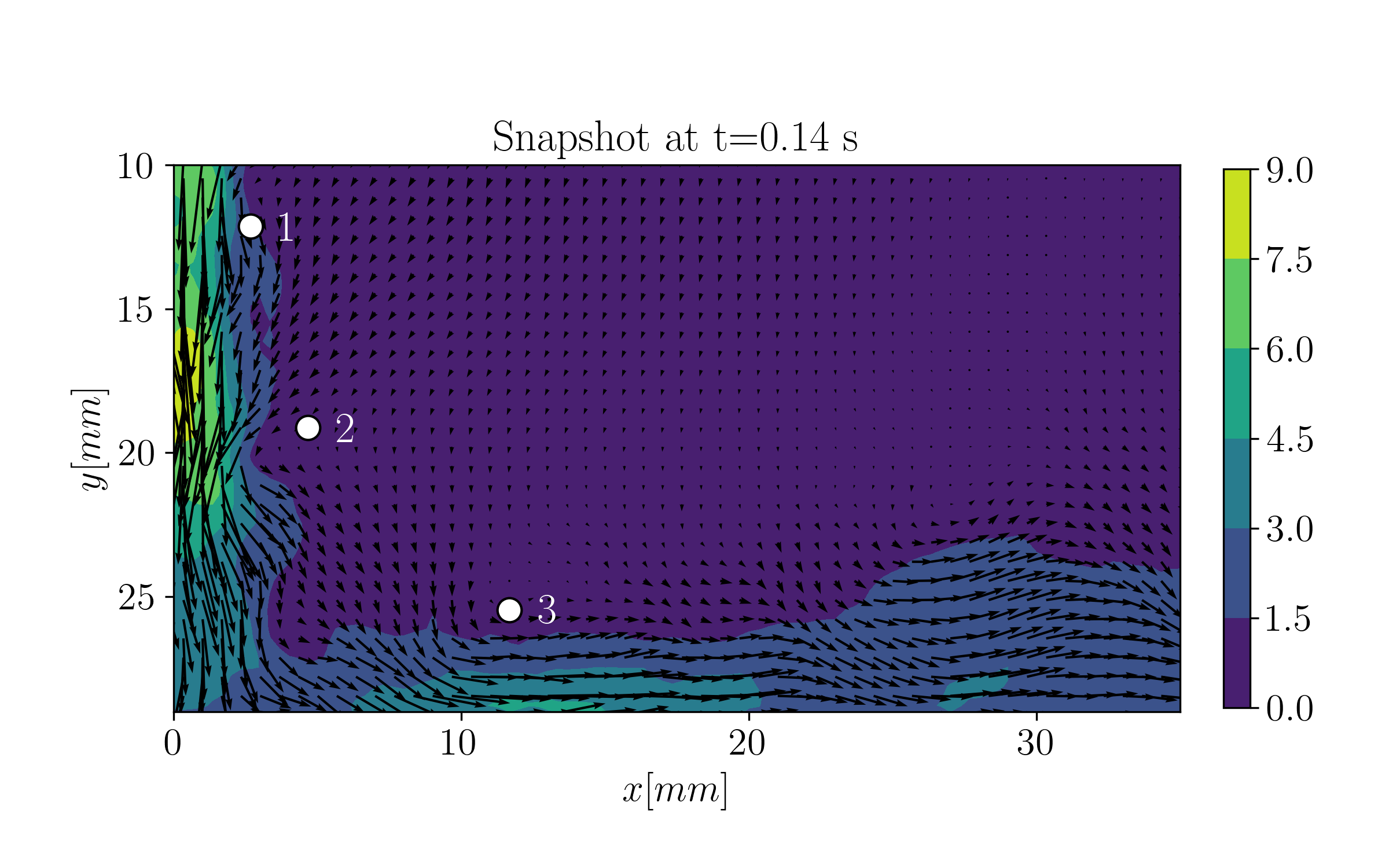}	}
	\end{subfigure}	    	\\
	\begin{subfigure}[\label{Jet_SP1}]{
			\includegraphics[width=.3\textwidth ]{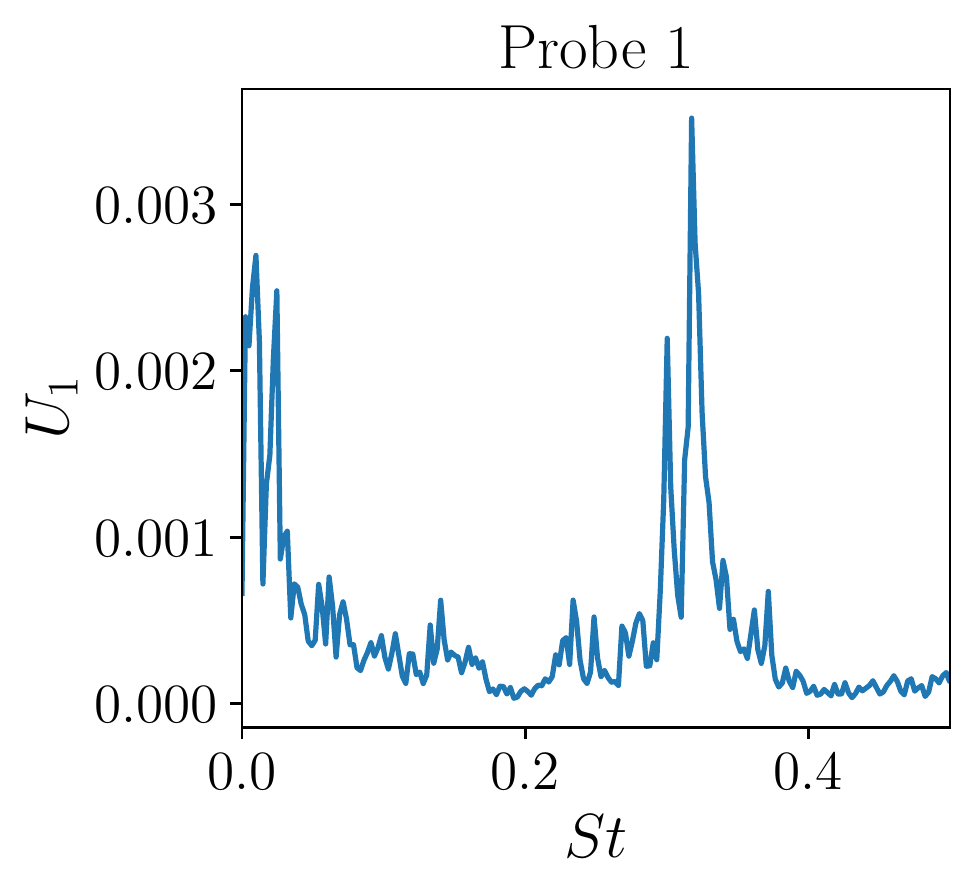}}
	\end{subfigure}	    	
	\begin{subfigure}[\label{Jet_SP2}]{
			\includegraphics[width=.3\textwidth ]{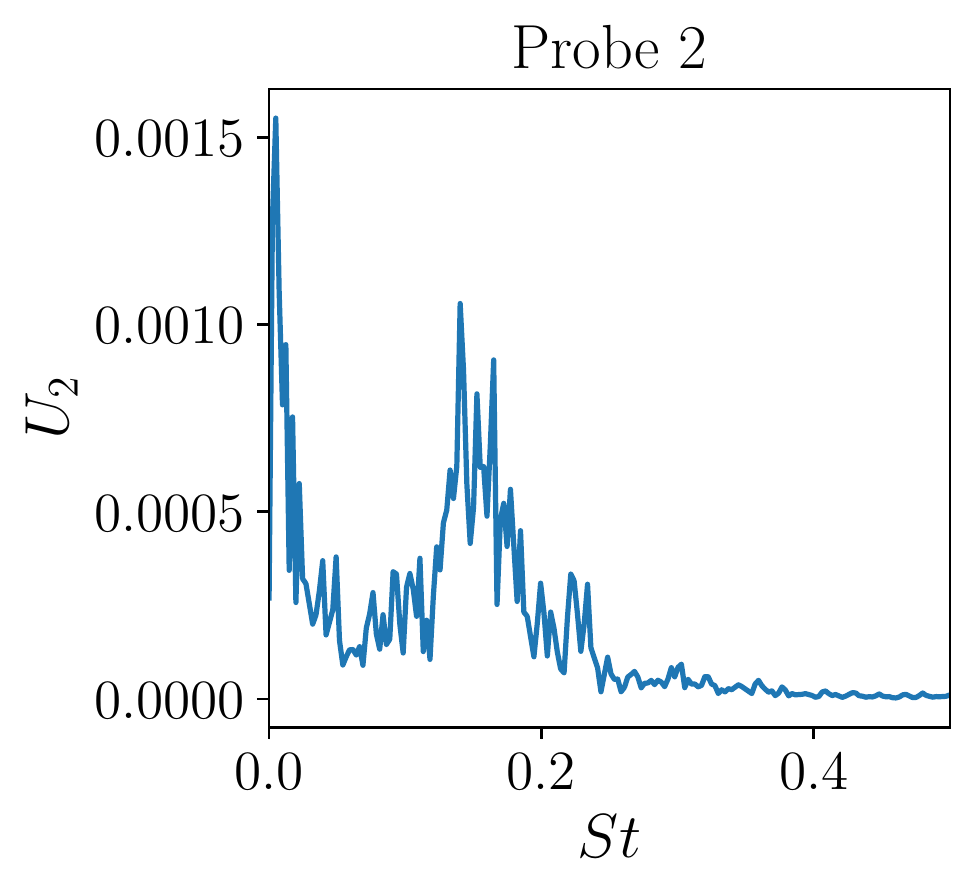}}
	\end{subfigure}
	\begin{subfigure}[\label{Jet_SP3}]{
			\includegraphics[width=.3\textwidth ]{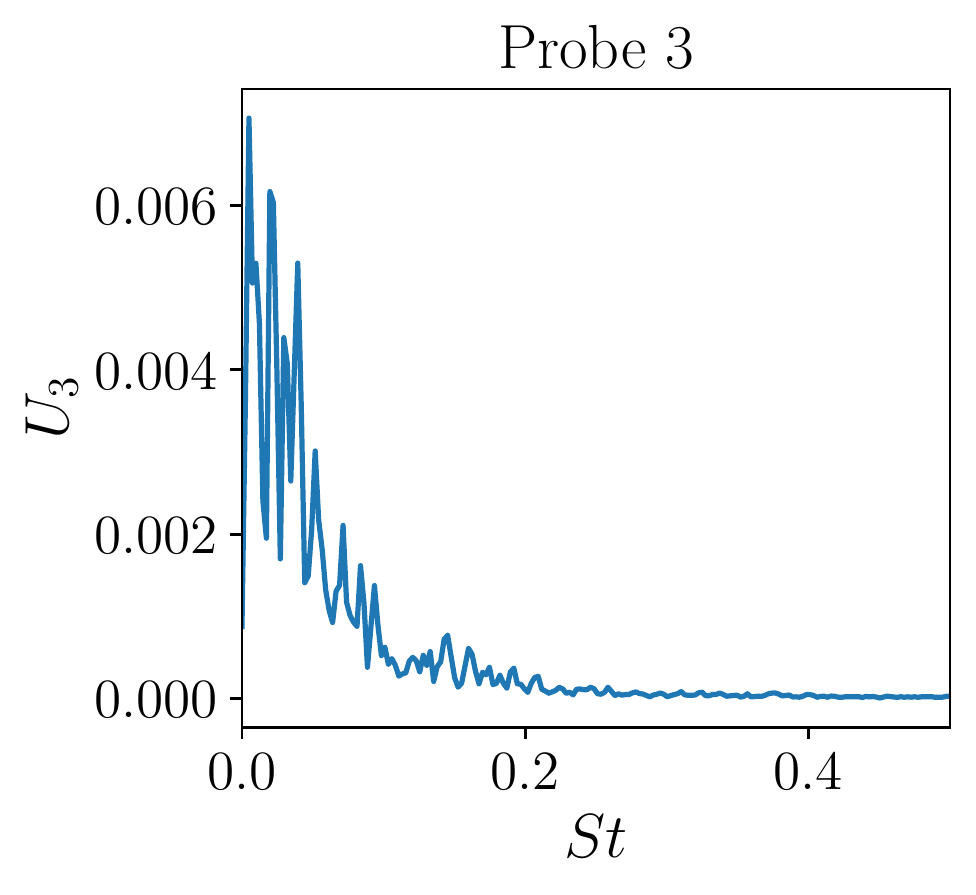}}
	\end{subfigure}
	
	\caption{a) Snapshot of the velocity field for the third test case: the TR-PIV measurements of an impinging gas jet. The figures b)-c)-d) show the power spectral density of the velocity magnitude in the probes 1, 2 and 3.} 
	\label{Original_Jet_RECO}
\end{figure}

              \subsection{Test Case 3: An impinging jet flow}\label{sec4p3}

              \begin{figure}[ht]
              	\centering
              	\begin{subfigure}[\label{CONV_PIV}]{
              			\includegraphics[width=.43\textwidth ]{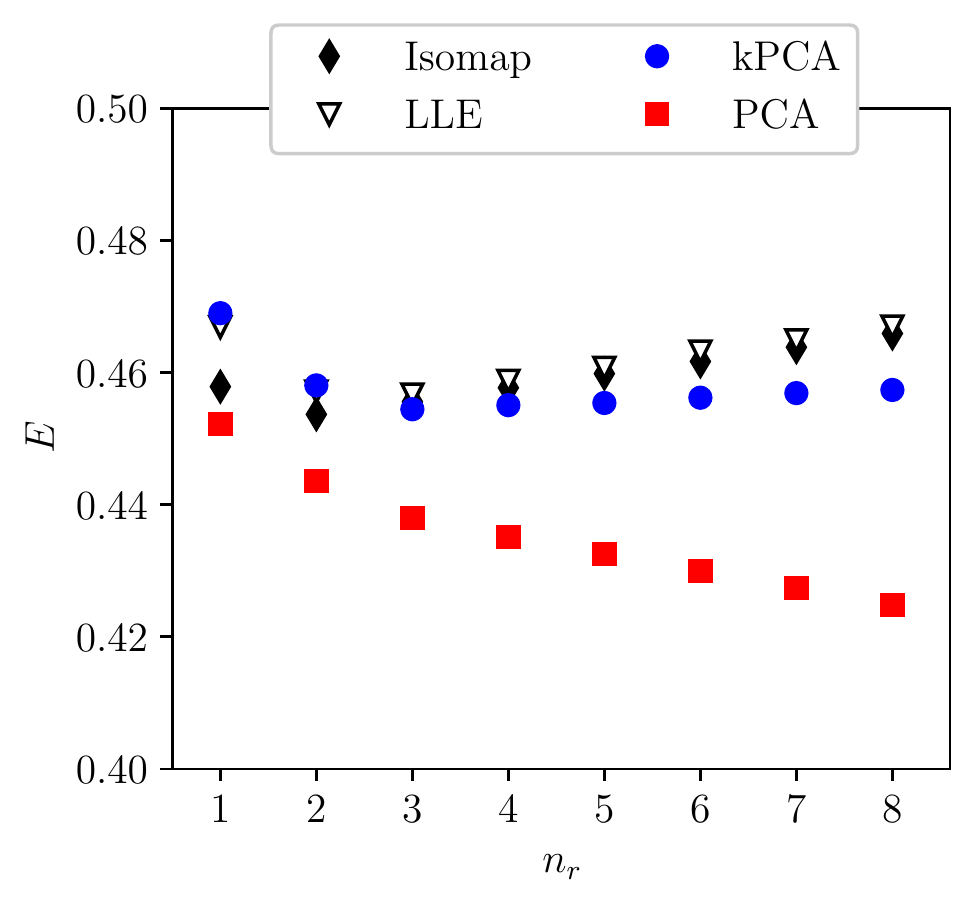}
              		}
              	\end{subfigure}
              	\begin{subfigure}[\label{R_V_PIV}]{
              			\includegraphics[width=.43\textwidth ]{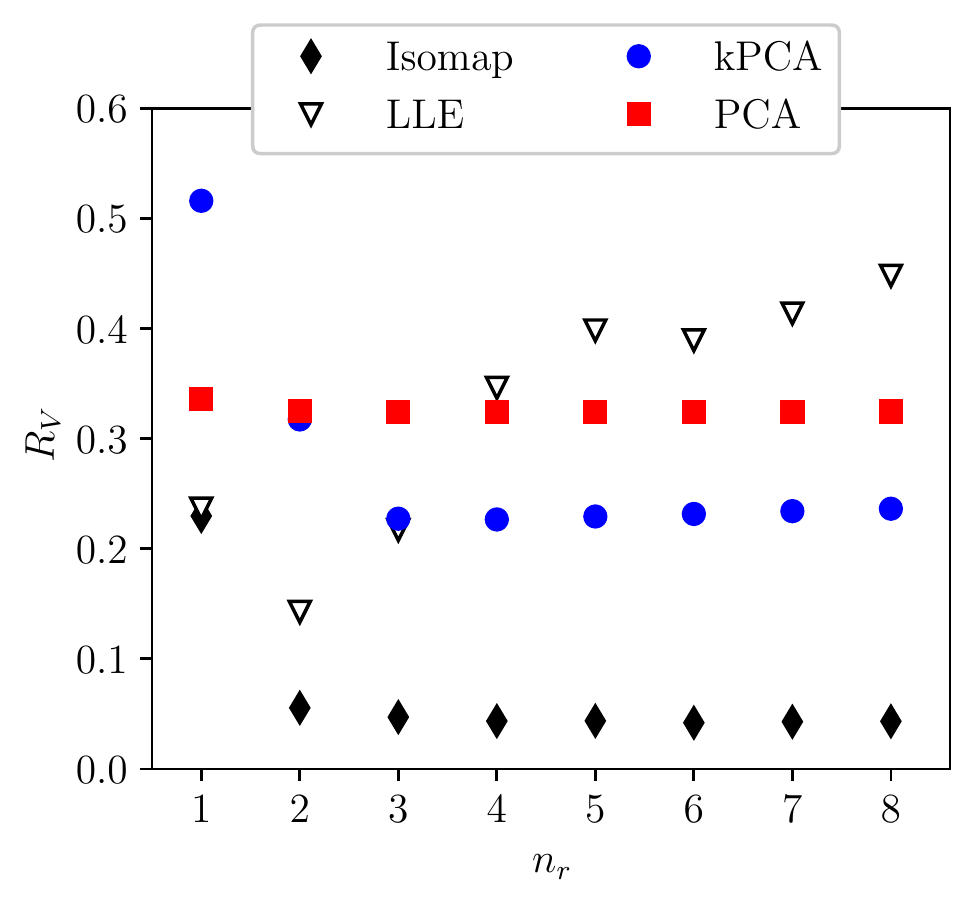}
              		}
              	\end{subfigure}	
              	\caption{Convergence of the relative $l_2$ error ($E$) and residual variance ($R_V$) as a function of the number of modes retained ($n_R$) for the test case 1, i.e. the set of PIV images.}
              	\label{Convergence_PIV_IM}
              \end{figure}
              
  We consider as third test case the a flow configuration which does not have an intrinsic low dimensionality like the previous. This is the flow of an impinging planar air jet, released at a mean velocity $U_J\approx 6.5$m/s from an opening of $H=4$mm at a distance of $Z=40$mm from a flat solid wall. This leads to a Reynolds number at the slot's exit of $Re_J=U_J H/\nu\approx 1700$. The dataset was obtained via time resolved PIV in \cite{Mendez2019} and consists of $n_t=2728$ fields with $60\times 114$ vectors, sampled at $f_s=2$kHz. The reader is referred to \cite{Mendez2019} for more details on the experimental set up and the related processing. The dataset has been made available at \url{https://osf.io/c28de/download}.
  
  A snapshot of the velocity field is shown in Figure \ref{Jet_SNAP}, together with the location of three probes. This configuration is inherently multiscale and was chosen as a benchmark for the multiscale POD (see also \cite{BarreiroVillaverde2021,Esposito2021,Ninni2020}).
              
 The Figures \ref{Jet_SP1}-\ref{Jet_SP3} show the power spectral density of the velocity magnitude at the three probes, scaled in terms of Strouhal number $St=f H/U_{J}$. The shear layer region close to the nozzle is characterized by the advection of roll-like vortex structures at $St\approx 0.3$ (see probe 1). The frequency of these vortices decreases as their velocity decreases while approaching the wall ($St\approx 0.17$ in probe 2). Downstream the stagnation point, in the wall jet region, the flow is much slower and characterized by large scale structures at $St<0.1$.

	  \section{Results}\label{sec:5}
	  
	 The following three subsections are dedicated to each of the three test investigated case.

		\subsection{Test Case 1: Background Removal Problem}\label{sec:5p1}
	
We first analyze the convergence of the implemented techniques for the first test case (cf. Section \ref{sec4p1}). The dataset is naturally in the range $[0,1]$, and no pre-processing is applied. It is worth recalling that the mean removal has a detrimental impact on the background removal problem, so neither the correlation matrix $\bm{K}$ nor the kernel matrix $\bm{K}_{\xi}$ are centered. On the other hand, the geodesic distance matrix in the ISOMAP was centered for numerical stability.

\begin{figure}[h!]
	\centering
	\includegraphics[width=.35\textwidth ]{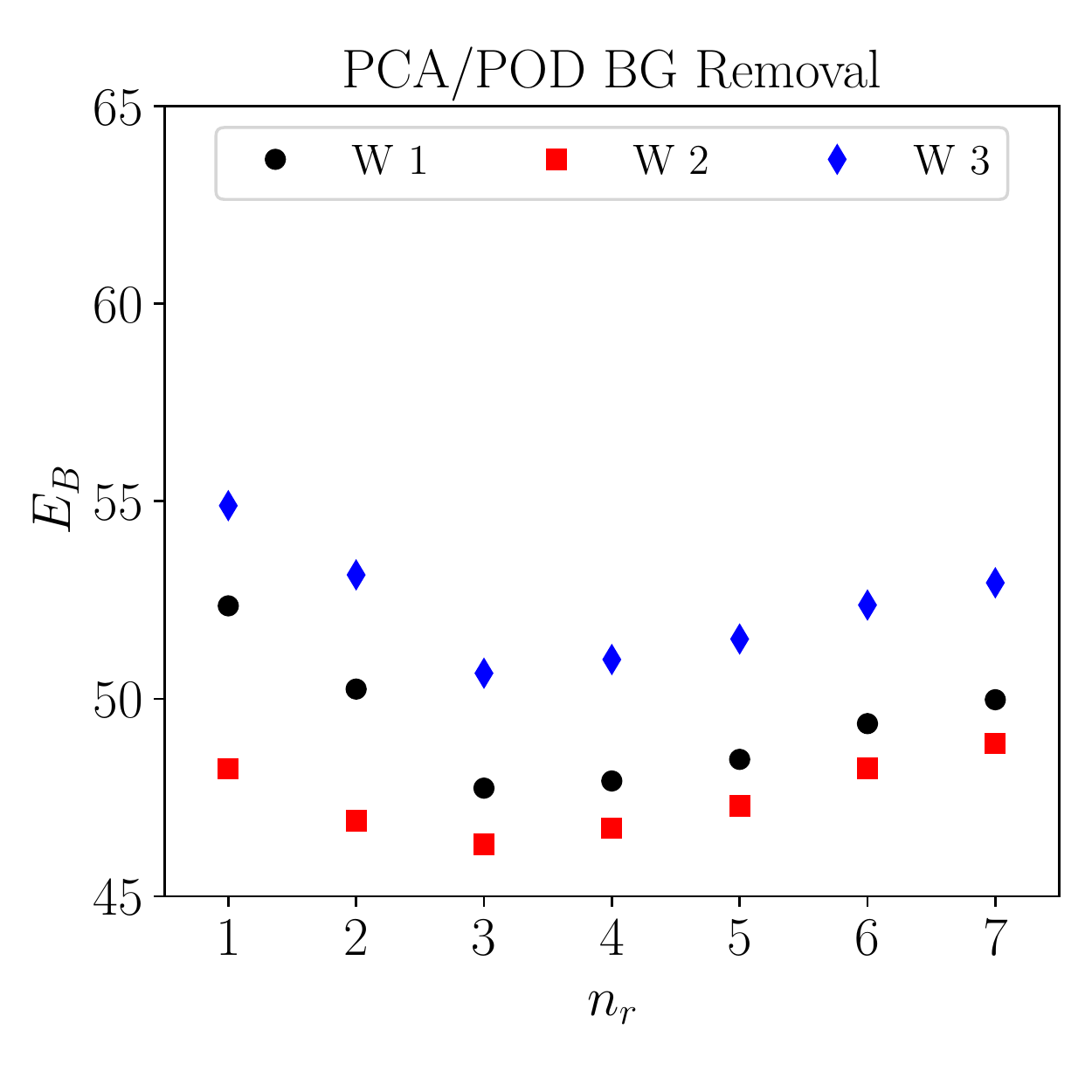}\\		
	\includegraphics[width=0.98\textwidth ]{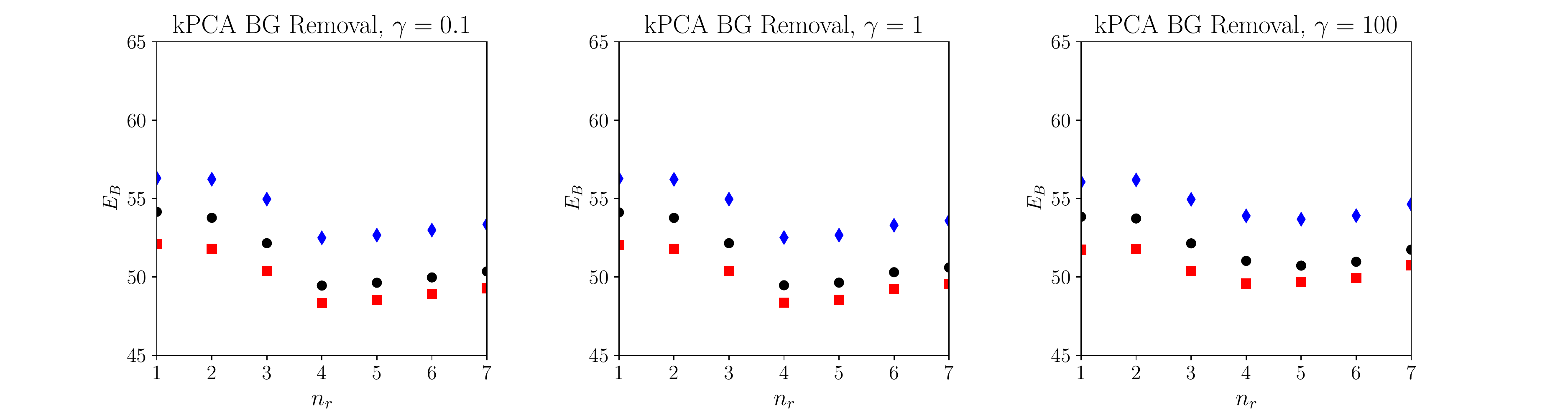}		
	\includegraphics[width=0.98\textwidth ]{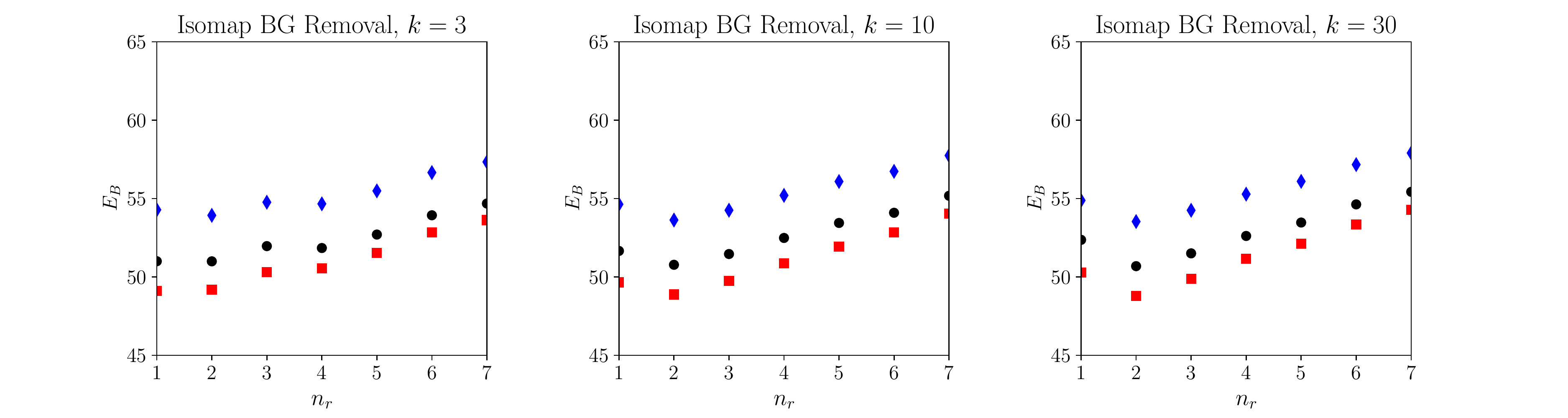}		
	\includegraphics[width=0.98\textwidth ]{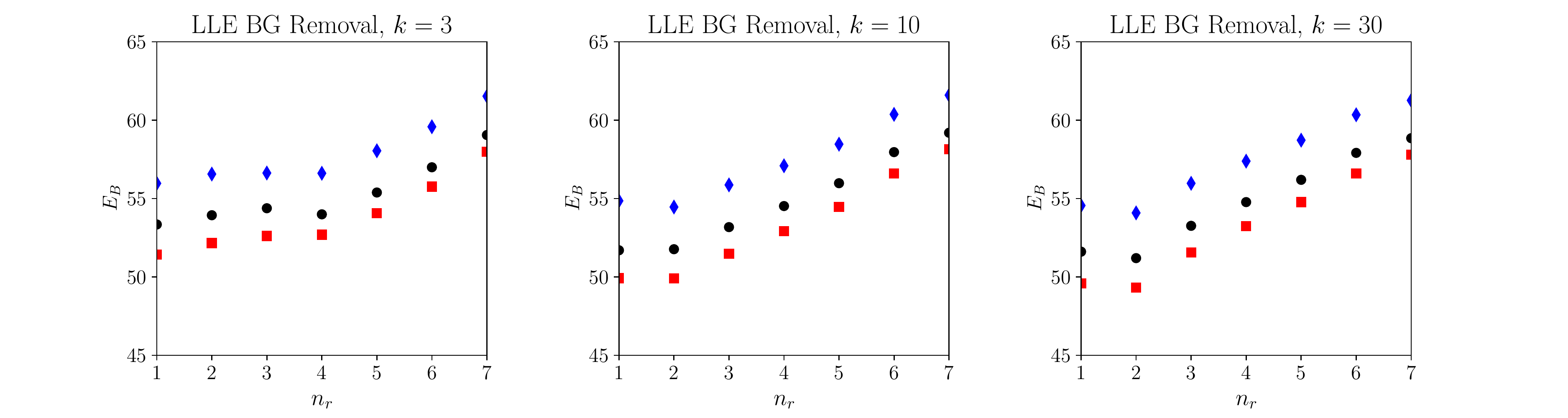}		
	\caption{Analysis of the error $E_B$ in \eqref{E_B} for the PCA kPCa, ISOMAPs and LLE in the three windows shown in Figure \ref{PIV_IMAGES_TEST_ZOOMS} and for various choices of $\gamma$ and $k$.} 
	\label{PIV_RES}
\end{figure}

The convergence results are shown in Figure \ref{Convergence_PIV_IM} in terms of relative $l_2$ errors on the left and residual variance $R_V$ on the right (see \eqref{ERR_l2} for definitions). Both quantities are ploted as function of $n_r$, that is the dimensionality of the encoding. The PCA has the best performance according to the first while the ISOMAPs has the best performance according to the second. These results were obtained with $\gamma=0.1$ for the kPCa and $k=10$ for the ISOMAPs and LLE.

\begin{figure}[h!]
	\centering	
	\includegraphics[width=0.98\textwidth ]{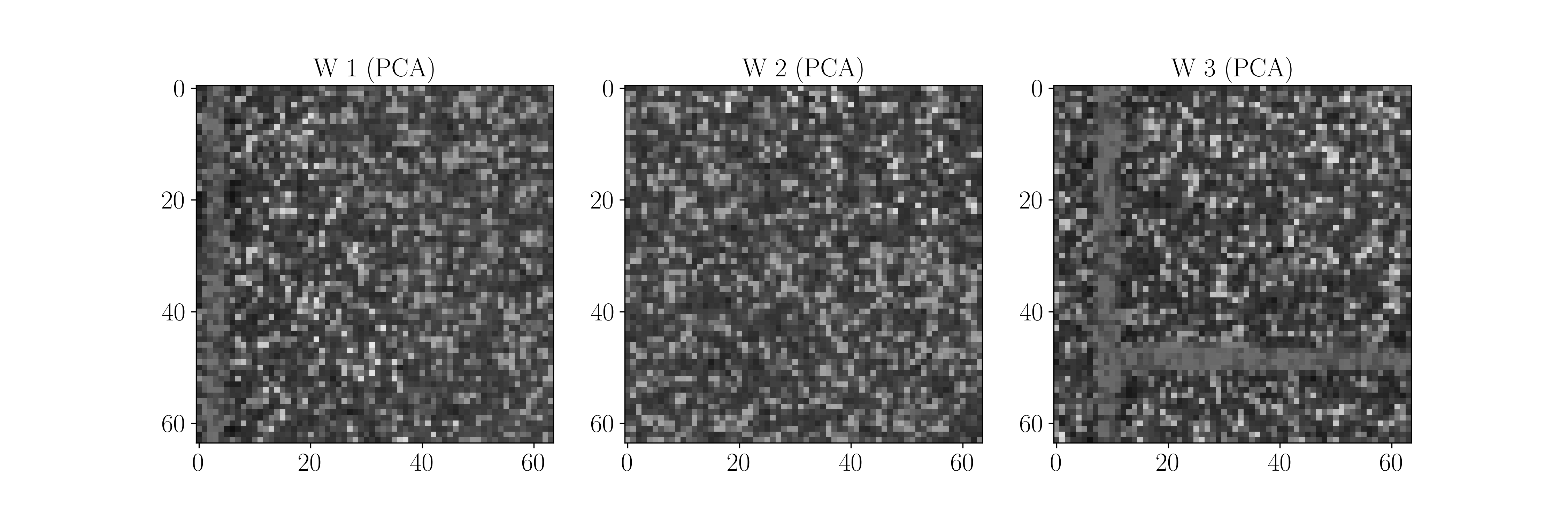}		
	\includegraphics[width=0.98\textwidth ]{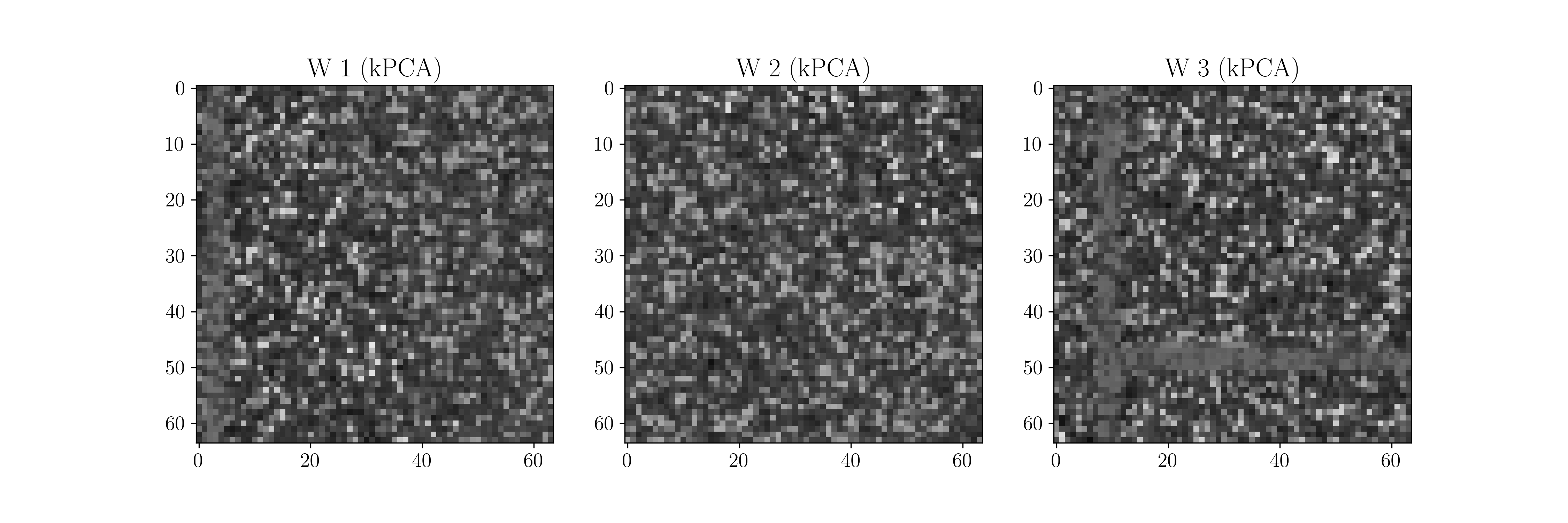}\\		
	\caption{Zoomed view on the windows W1, W2 and W3 (see Figure \ref{PIV_IMAGES_TEST_ZOOMS}) after background removal with PCA (top) and kPCA (bottom). The results with ISOMAPs and LLE are indistinguishable from the ones with the kPCA and are thus omitted.} 
	\label{PIV_RES_HPO}
\end{figure}

Concerning the $l_2$ error convergence, the slow decay is expected for the set of PIV images since most of the variance is due to the contribution of the PIV particles, which is random. Interestingly, at $n_r>2$ the error increases with $n_r$ for the nonlinear techniques. While this might be due to limits of the simple implemented decoder, this result is particularly interesting when considering the performances in terms of residual variance $R_V$ (Figure \ref{R_V_PIV}): this is monotonically decreasing behaviour for the ISOMAP but quickly saturates for the PCA. This shows that minimizing $E$ and minimizing $R_V$ are contrasting goals for the problem at hand. The performances of kPCA and LLE fall between the PCA and ISOMAPS, with the LLE out-performing the kPCA in theresidual variance for $n_r<4$.

In this filtering problem, it is interesting to have the approximated images as close as possible to the underlying ideal PIV sequence. We thus analyze the performances of these methods in terms of $E_B$ (see \eqref{E_B} for definition). Figure \ref{PIV_RES} collects the results for all methods and the three windows in Figure \ref{PIV_IMAGES_TEST_ZOOMS} and different values for $\gamma$ and $k$ in the nonlinear techniques.

The figure on the top is related to the PCA, acting as a reference. As expected, the background noise can be removed more easily in window W2 than in window W3. For the PCA, the best performances are achieved using $n_r=3$, at which all curves have a minimum. This minimum is due to the convergence of the decomposition (see also Figure \ref{Convergence_PIV_IM}): the lower the reconstruction error, the more the approximation resembles the provided images, including background noise \emph{and} particles. In other words, at larger $n_r$, the PCA background removal becomes too aggressive, and the resulting erosion of the PIV particles outweighs the gain in background removal.

A similar trend is observed in kPCA, which also has a minimum $E_B$ at $n_r=4$. This corresponds (see \ref{Convergence_PIV_IM}) to a region of rather flat convergence for both $E$ and $R_V$. Interestingly, the value of $\gamma$ does not significantly impact the performances within the investigated four orders of magnitude ($\gamma=0.1$ to $\gamma=100$). Finally, the performances of ISOMAPs and LLE are similar, with the first slightly better than the second. The parameter $k$ has a moderate impact. For $k=10$ and $k=30$, a minima appears at $n_r=2$ and the error increases linearly for $n_r>2$. 

Taking the best configuration for each method ($n_r=3$ for the PCA, $n_r=3$ and $\gamma=0.1$ for the kPCA and $n_r=2$ and $k=10$ for ISOMAPs and LLE) produced nearly identical results. The filtered results (to be contrasted with Figure 
\ref{PIV_IMAGES_TEST_ZOOMS}) are shown in Figure \ref{PIV_RES_HPO}. The results from LLE and ISOMAPs are omitted because these are indistinguishable from the results obtained via kPCA. All methods remove three of the main sources of background noise but suffer with the first (the vertical and horizontal stripes) because this saturates in various snapshots. Better removal of these stripes can be obtained if $n_r=1$ but at the cost of worsening the performances far from these regions. 

Considering that this test case is extremely challenging from a denoising point of view, these results are largely satisfactory regarding background removal. On the other hand, these results show that none of the implemented nonlinear approaches outperforms the classic PCA in the investigated background removal problem. The better performances in terms of residual variance (hence manifold mapping) are of no help in this test case.

	\begin{figure}[h!]
		\centering
		\begin{subfigure}[\label{CONV_CYL}]{
				\includegraphics[width=.43\textwidth ]{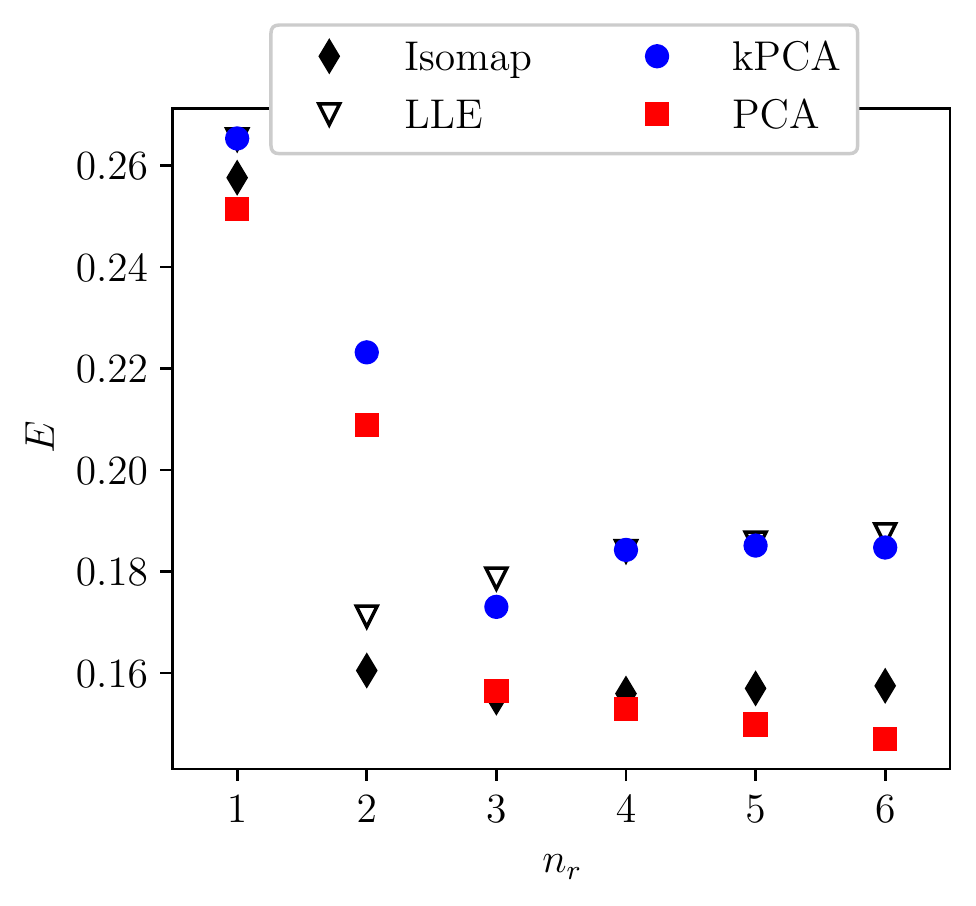}
			}
		\end{subfigure}
		\begin{subfigure}[\label{R_V_CYL}]{
				\includegraphics[width=.43\textwidth ]{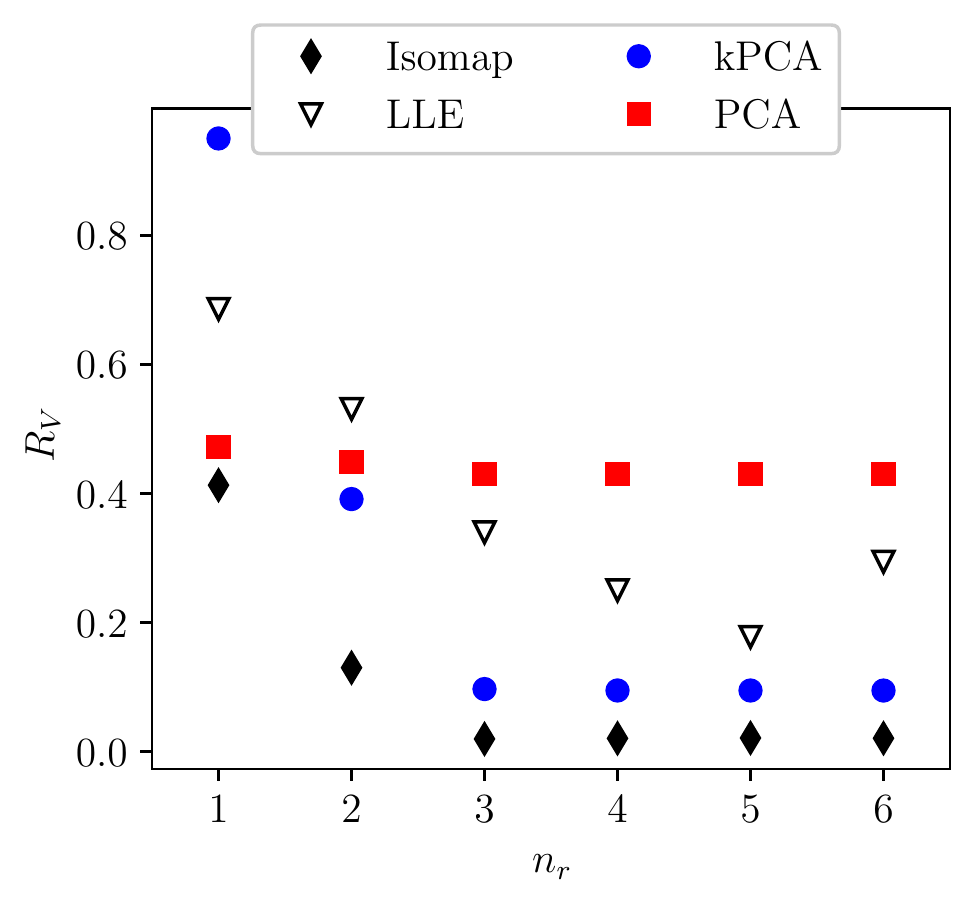}
			}
		\end{subfigure}	
		\caption{Same as figure \ref{Convergence_PIV_IM}, but for the TR-PIV fields of the cylinder wake flow (Test Case 2, Section \ref{sec4p2}).}
		\label{Convergence_CYL}
	\end{figure}		
	
\subsection{Test Case 2: Transient Flow past a Cylinder}\label{sec5p2}

We here consider the second test case (cf. Section \ref{sec4p2}). The dataset was normalized in the range $[0,1]$. No mean removal was performed since the flow is in transient conditions. Figure \ref{Convergence_CYL} shows the convergence of the $l_2$ norm error (Fig. \ref{CONV_CYL}) and the residual variance (Fig. \ref{R_V_CYL}) for all the implemented methods. For the kPCA, the parameter $\gamma$ is computed by setting $\kappa_m=0.001$ (see Section \ref{sec:3p2}) while in the ISOMAPS and LLE we have $k=10$. The same number of nearest neighbours is used in the linear decoder for the nonlinear methods. These parameters were selected using a grid search for hyperparameter tuning, setting the minimization of $E_B(n_r)$ at $n_r=3$ as a goal.

The results in Figure \ref{Convergence_CYL} confirm that the minimization of $E_B(n_r)$ does not imply the minimization of $R_V(n_r)$ (see Sec \ref{sec:3p5}), as also observed in the previous section. However, in this problem, the ISOMAPs outperform all decompositions in both metrics. The kPCA gives the second best performance in terms of $R_V$ convergence while the PCA reaches a plateau already at $n_r>2$. As discussed in Section \ref{sec4p2}, this test case is characterized by an intrinsically low dimensionality, with $n_r=3$ known to capture the essential features (see also \cite{Mendez2020}). It is thus interesting that both kPCA and ISOMAps feature an `elbow' at $n_r=3$ in the convergence of the residual variance error. 

\begin{figure}[h!]
	\centering	
	PCA (POD)		
	\begin{minipage}{0.4\textwidth}
		\centering
		\includegraphics[trim={0cm 0.3cm 0 0.2cm},clip, width=0.97\linewidth]{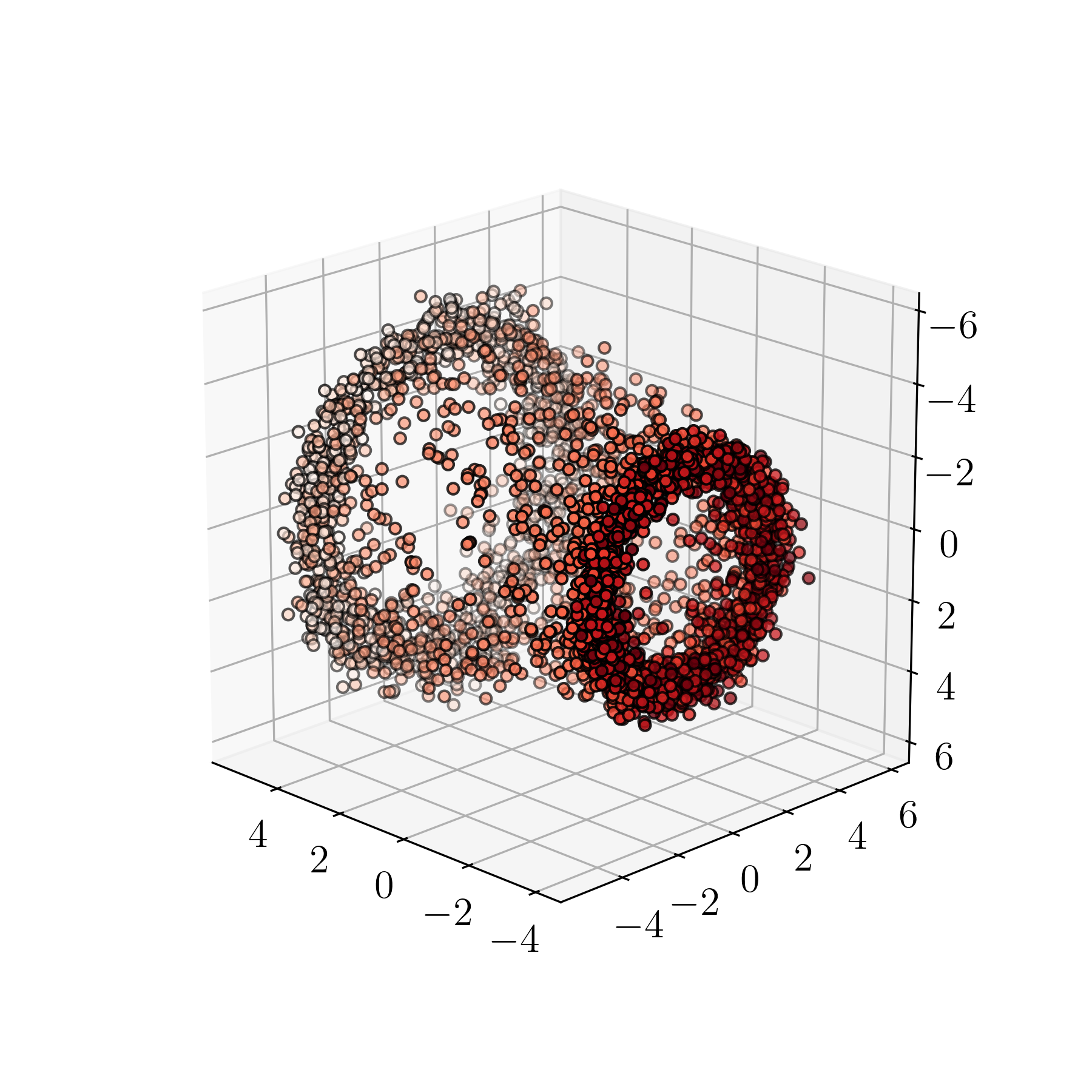}
	\end{minipage}			
	\begin{minipage}{0.4\textwidth}
		\centering
		\vspace{0.2mm}
		\includegraphics[width=0.95\linewidth]{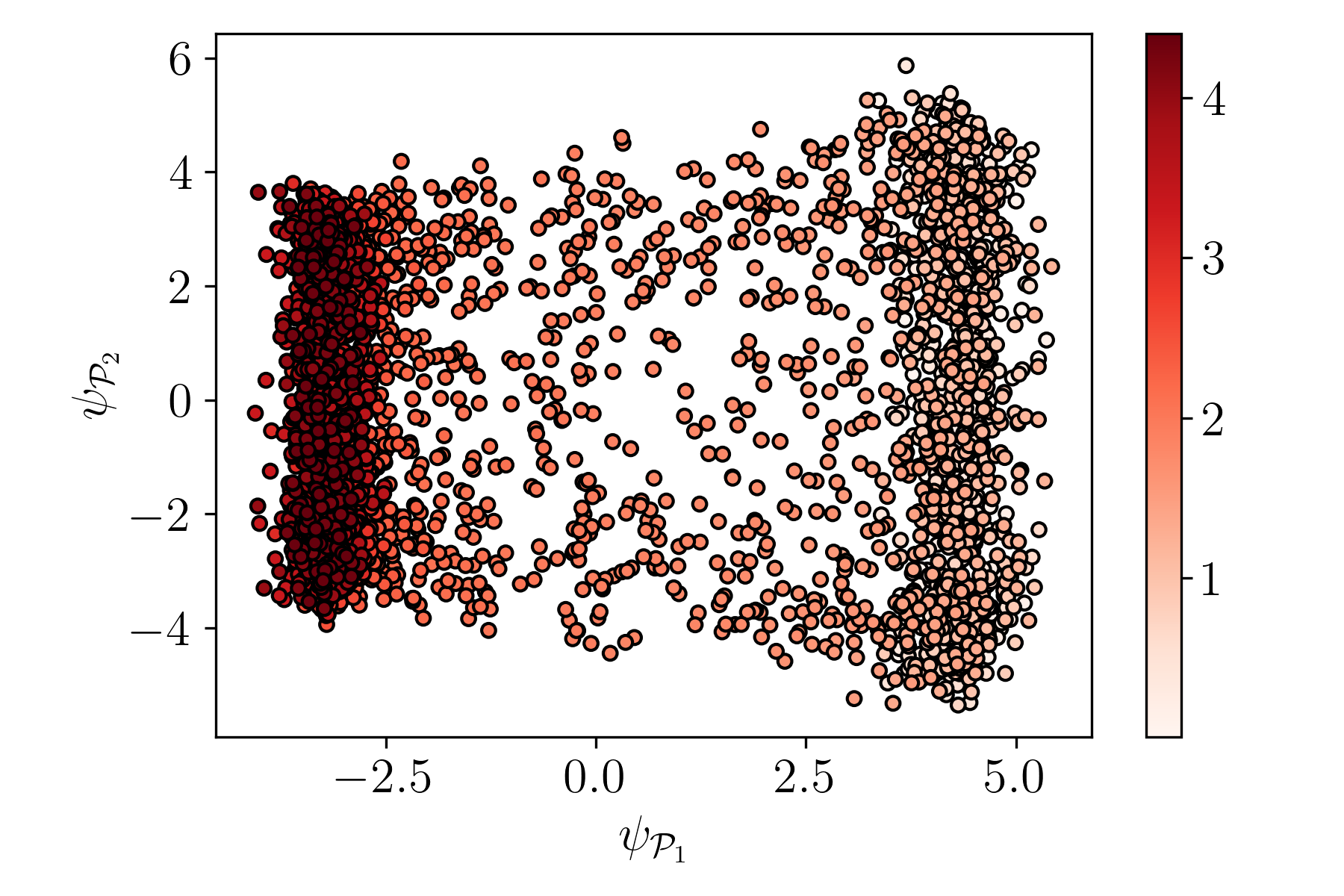}		\end{minipage}\\
	\vspace{-5mm}
	kPCA
	\begin{minipage}{0.4\textwidth}
		\centering
		\includegraphics[trim={0cm 0.3cm 0 0.2cm},clip,width=0.95\linewidth]{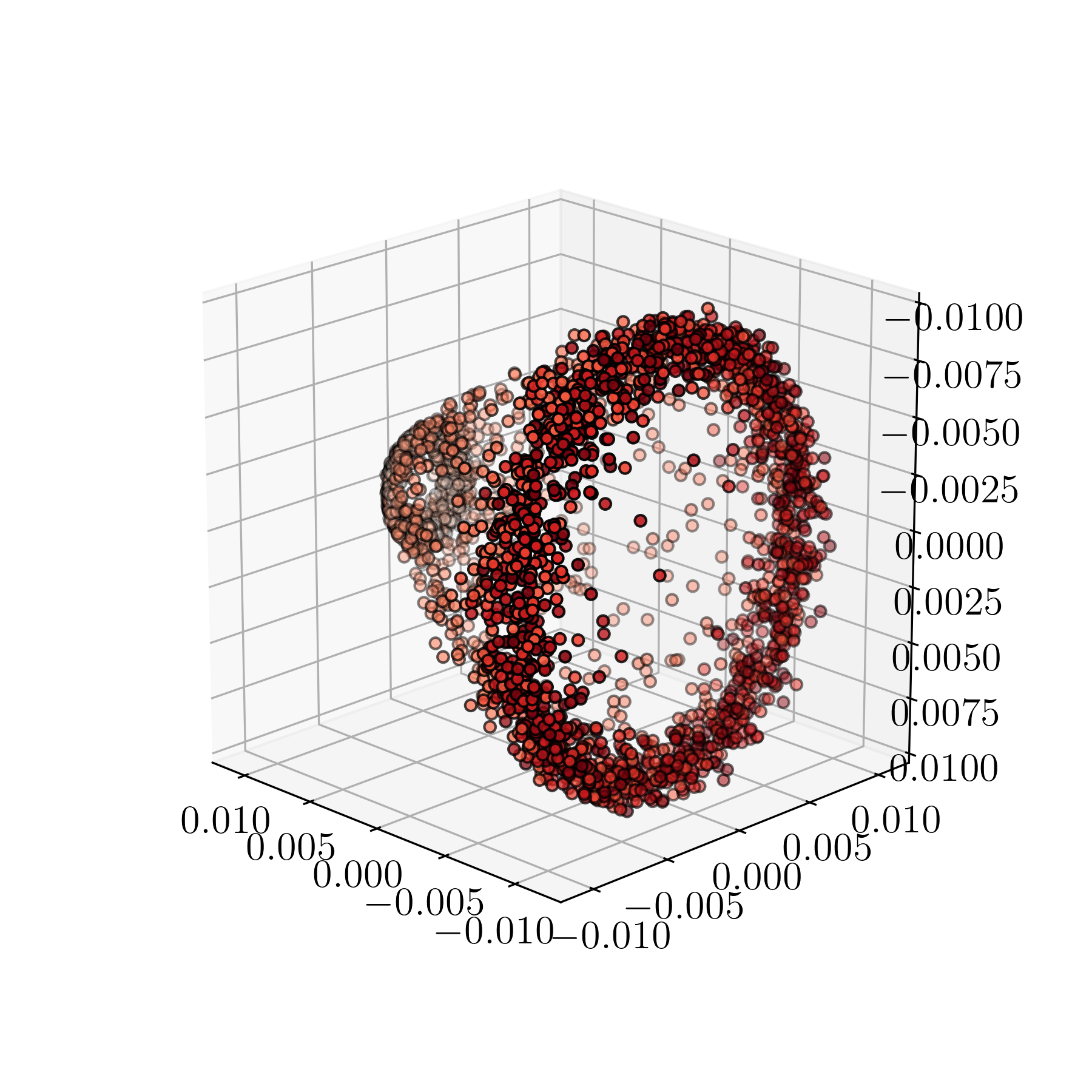}
	\end{minipage}			
	\begin{minipage}{0.4\textwidth}
		\centering
		\vspace{0.2mm}
		\includegraphics[width=0.95\linewidth]{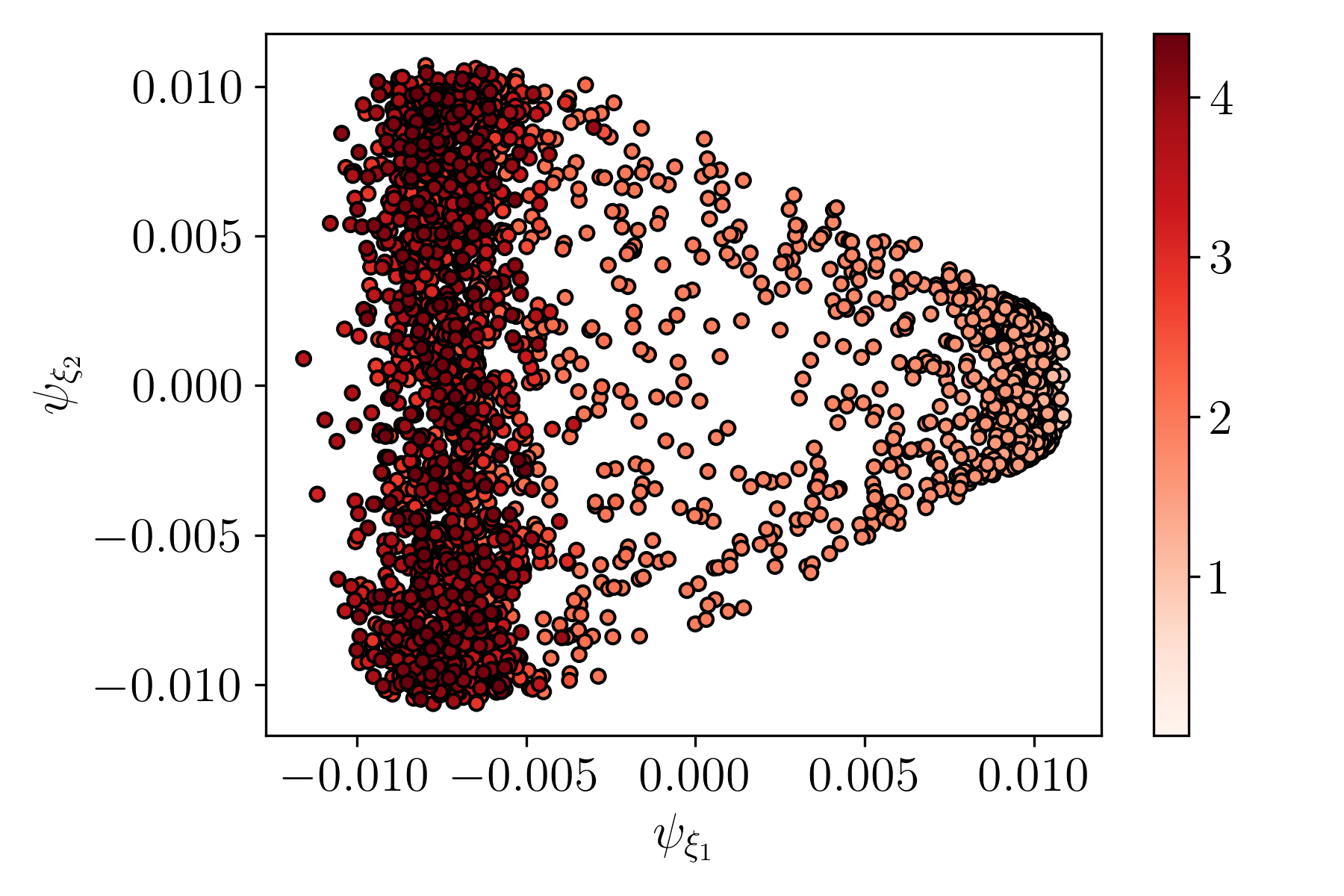}		
	\end{minipage}\\
	\vspace{-5mm}
	ISOMAPs
	\begin{minipage}{0.4\textwidth}
		\centering
		\includegraphics[trim={0cm 0.3cm 0 0.2cm},clip,width=0.95\linewidth]{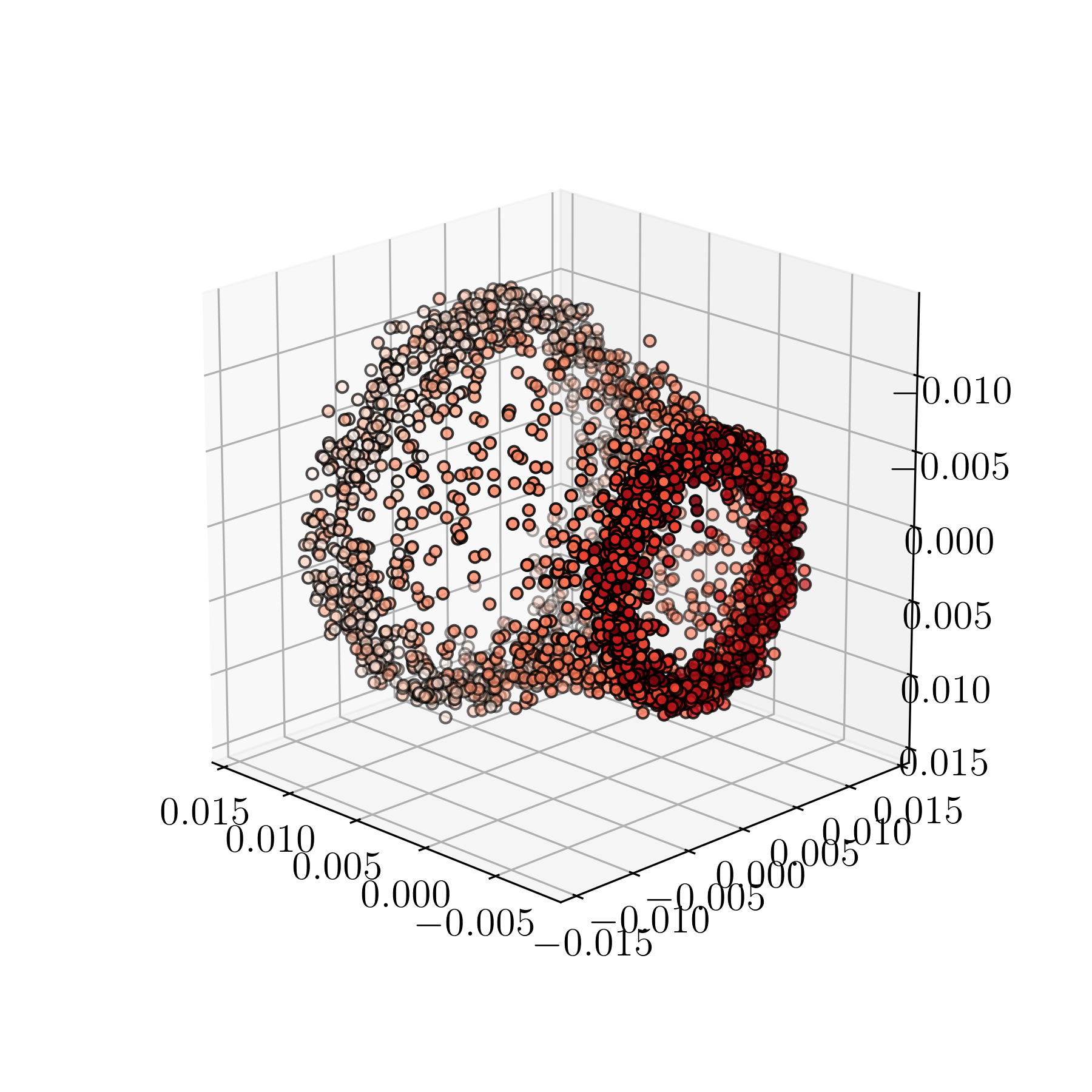}
	\end{minipage}			
	\begin{minipage}{0.4\textwidth}
		\centering
		\vspace{0.2mm}
		\includegraphics[width=0.95\linewidth]{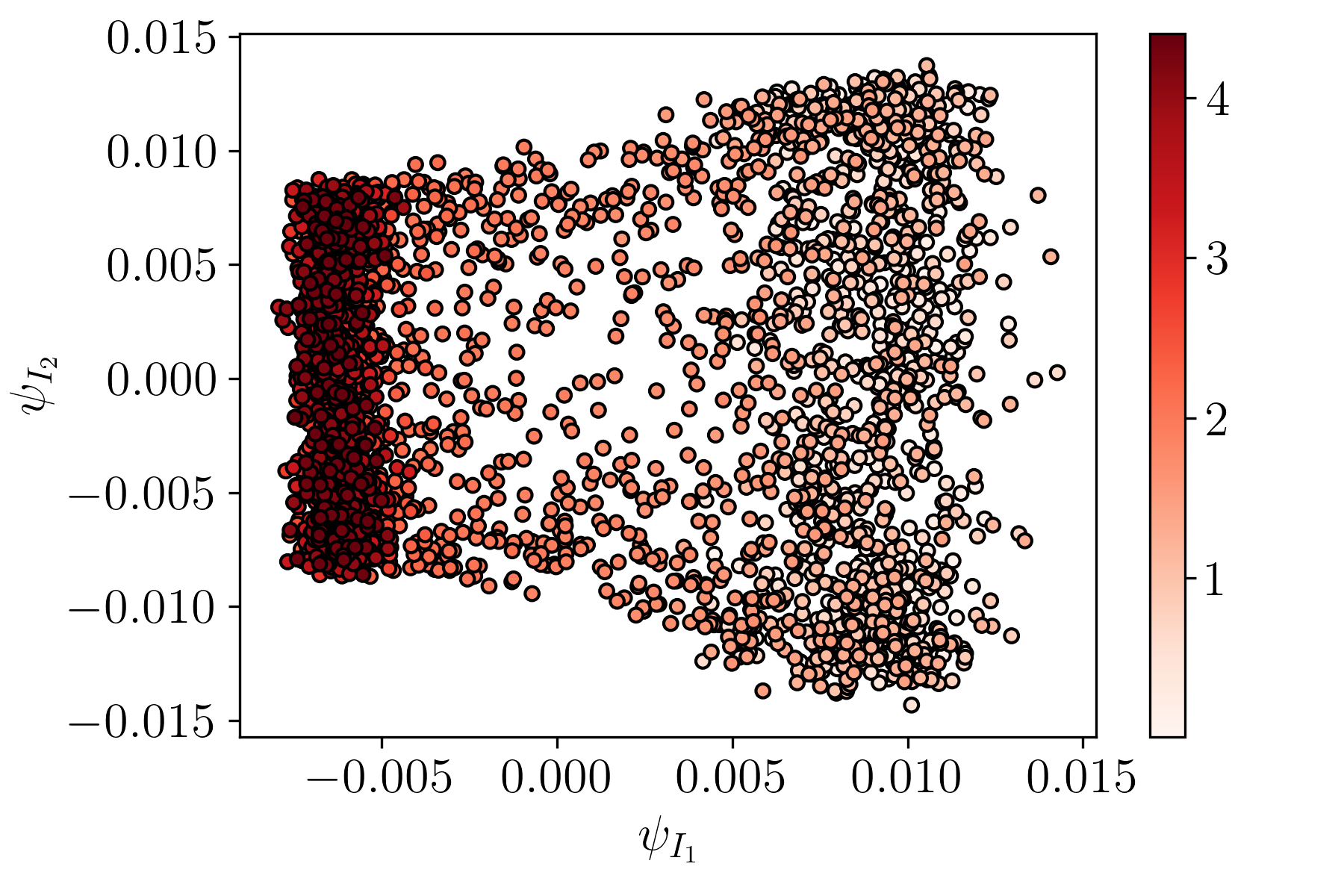}		
	\end{minipage}\\
	\vspace{-5mm}
	LLE
	\begin{minipage}{0.4\textwidth}
		\centering
		\includegraphics[trim={0cm 0.3cm 0 0.2cm},clip,width=0.95\linewidth]{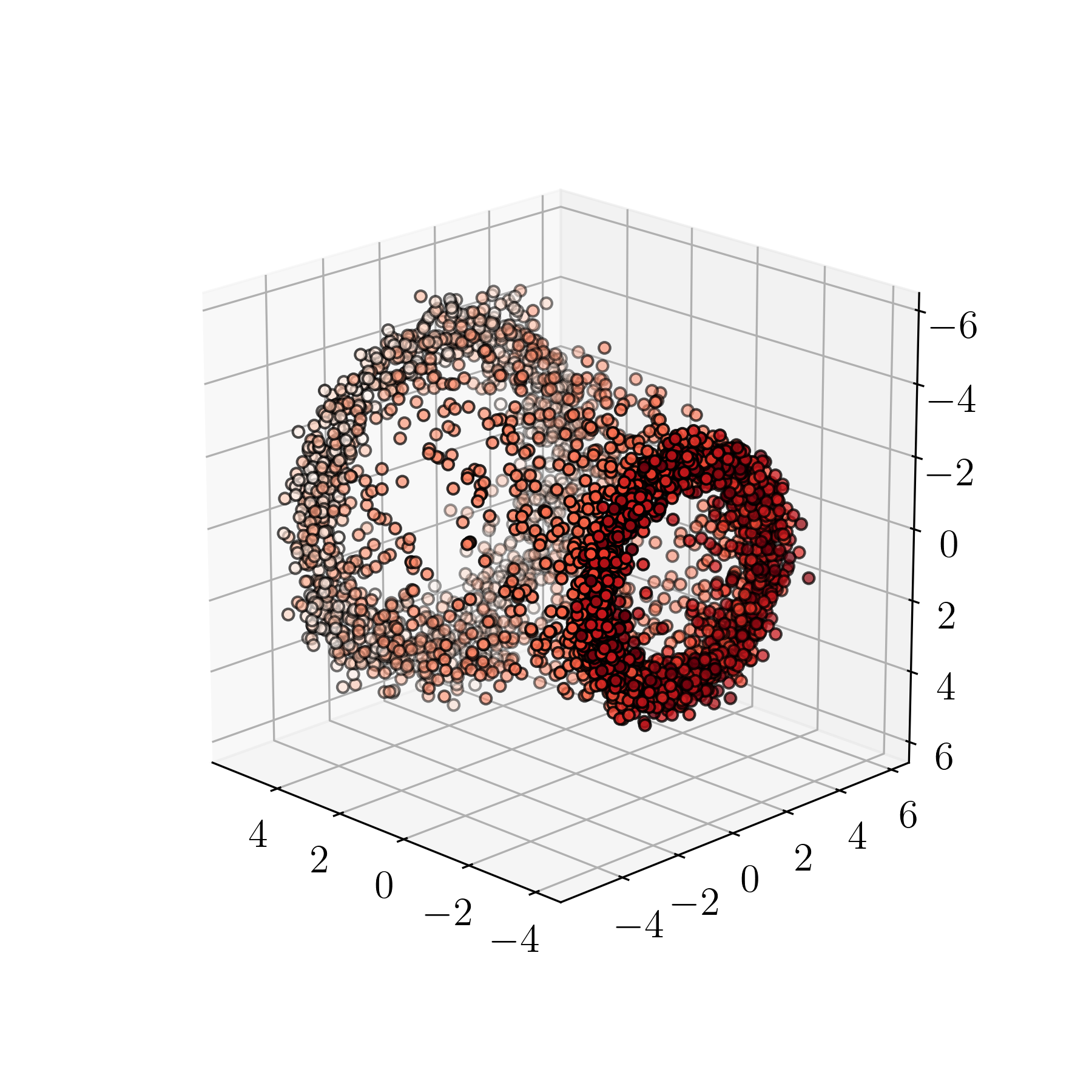}
	\end{minipage}			
	\begin{minipage}{0.4\textwidth}
		\centering
		\vspace{0.2mm}
		\includegraphics[width=0.95\linewidth]{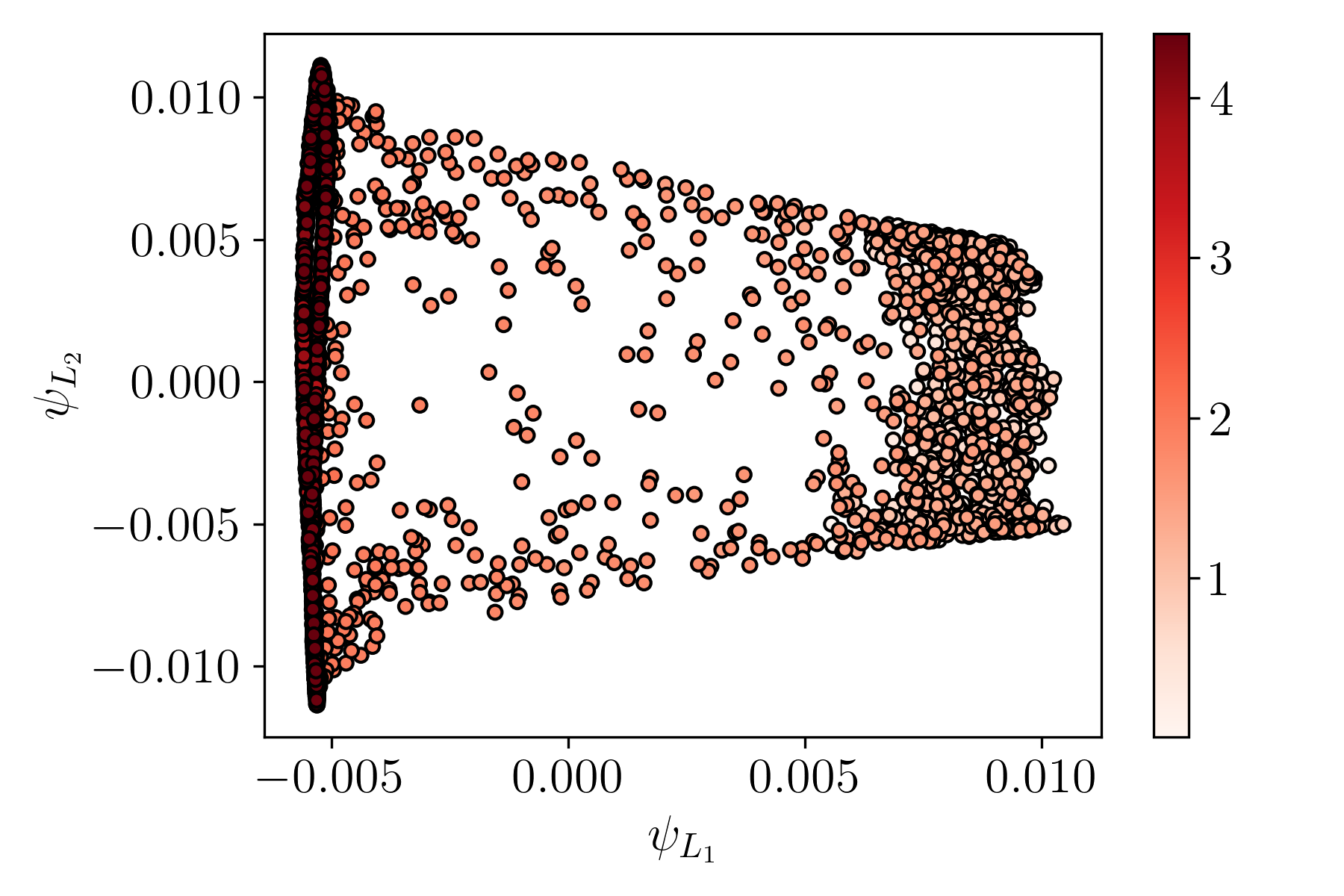}		
	\end{minipage}\\
	\vspace{-5mm}

	\caption{Trajectory of in low dimensional representation of the transient turbulent cylinder flow (test case 2, described in section \ref{sec5p2}) with $n_r=3$ for the PCA(POD), kPCA, ISOMAPs and LLE. The figures on the left show the 3D view; the figures on the right show the `lateral' projection to highlight the system evolution during the transient.}
	\label{CYL_TRAJS}
\end{figure}

The evolution of the system in the reduced space ($\mathbb{R}^3$) is shown in Figure \ref{CYL_TRAJS} for all investigated techniques. The figures on the left show a 3D view, while the figure on the right shows the projection on the $\psi_2-\psi_3$ plane, which clearly shows the evolution of the transient. In all figures, the markers are colored by the time axes (in seconds) to reveal the system trajectories (cf. Figure \ref{CYL_TEST_CASE}b). All reduced representations picture the vortex shedding as a circular orbit in the $\psi_2-\psi_3$ plane. As time evolves and the system moves from one steady state to another, an axisymmetric trajectory connects the two orbits.

\begin{figure}[h!]
	\centering  
	\begin{subfigure}[\label{Snap_Cyl_PCA}]{
			\includegraphics[trim={0cm 0.5cm 1.5cm 0.5cm},clip,width=.47\textwidth ]{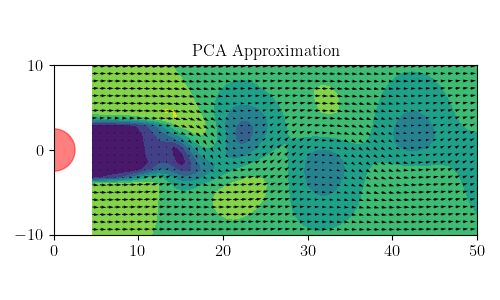}}	
	\end{subfigure}	    	
	\begin{subfigure}[\label{Snap_Cyl_kPCA}]{
			\includegraphics[trim={0cm 0.3cm 1cm 0.25cm},clip,width=.47\textwidth ]{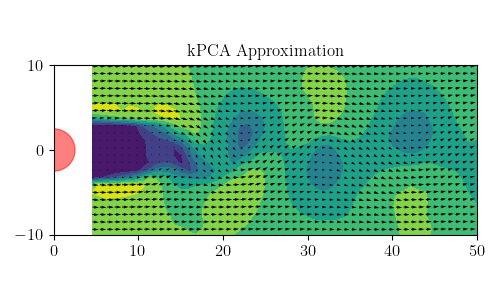}}
	\end{subfigure}	\\
	\begin{subfigure}[\label{Snap_Cyl_ISO}]{
			\includegraphics[trim={0cm 0.3cm 0.5cm 0.25cm},clip,width=.47\textwidth ]{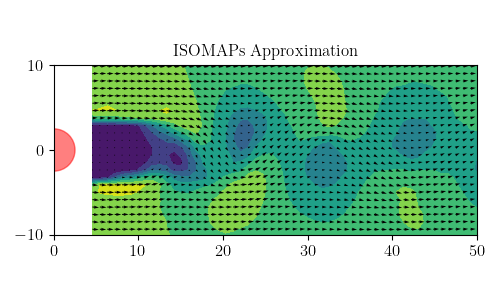}}
	\end{subfigure}	
	\begin{subfigure}[\label{Snap_Cyl_LLE}]{
			\includegraphics[trim={0cm 0.3cm 0.5cm 0.25cm},clip,width=.47\textwidth ]{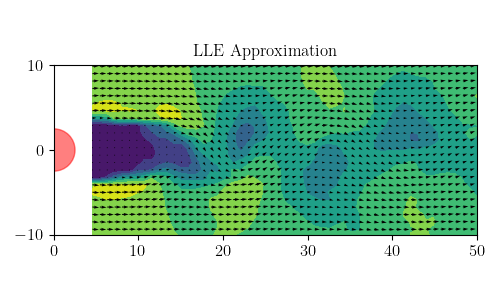}}
	\end{subfigure}	
	\caption{Flow field reconstruction with $n_r=3$ for the four decomposition. The selected snapshot is the one for which the original data is plotted in Figure \ref{Snap_Cyl}.} 
	\label{CYL_RECONSTRUCTIONS}
\end{figure}

The manifold identified by the PCA appears as a truncated cone, with the shedding at higher speed (earliest times) associated with larger diameters' orbit. This is similar for ISOMAPs, although the evolution during the transient is curved and more closely follows the smooth evolution of the free stream. In kPCA and LLE, the manifold is also close to a truncated cone but with the opposite position of the orbits (larger orbit at the slowest velocity ). In the case of kPCA, the slope in the cone is much amplified compared to PCA. Although these differences might appear minor, it is worth stressing that they are linked to significantly different values of $E_B$ and $R_V$ (cf. Figure \ref{Convergence_CYL}). To qualitatively appreciated the quality of the reconstructions, Figure \ref{CYL_RECONSTRUCTIONS} shows one snapshot of the flow field using approximation using $n_r=3$ for the four approaches. The selected snapshot is the one for which the original data is shown in Figure \ref{Snap_Cyl}. All reconstructions are remarkably close to the original snapshot. The PCA reconstruction provides the smoothest fields, blurring out fine details of the vortical structures, while the nonlinear techniques recover finer flow details more closely.

\FloatBarrier

\subsection{Test Case 3: Impinging Jet Flow}

\begin{figure}[h!]
	\centering
	\begin{subfigure}[\label{CONV_JET}]{
			\includegraphics[width=.45\textwidth ]{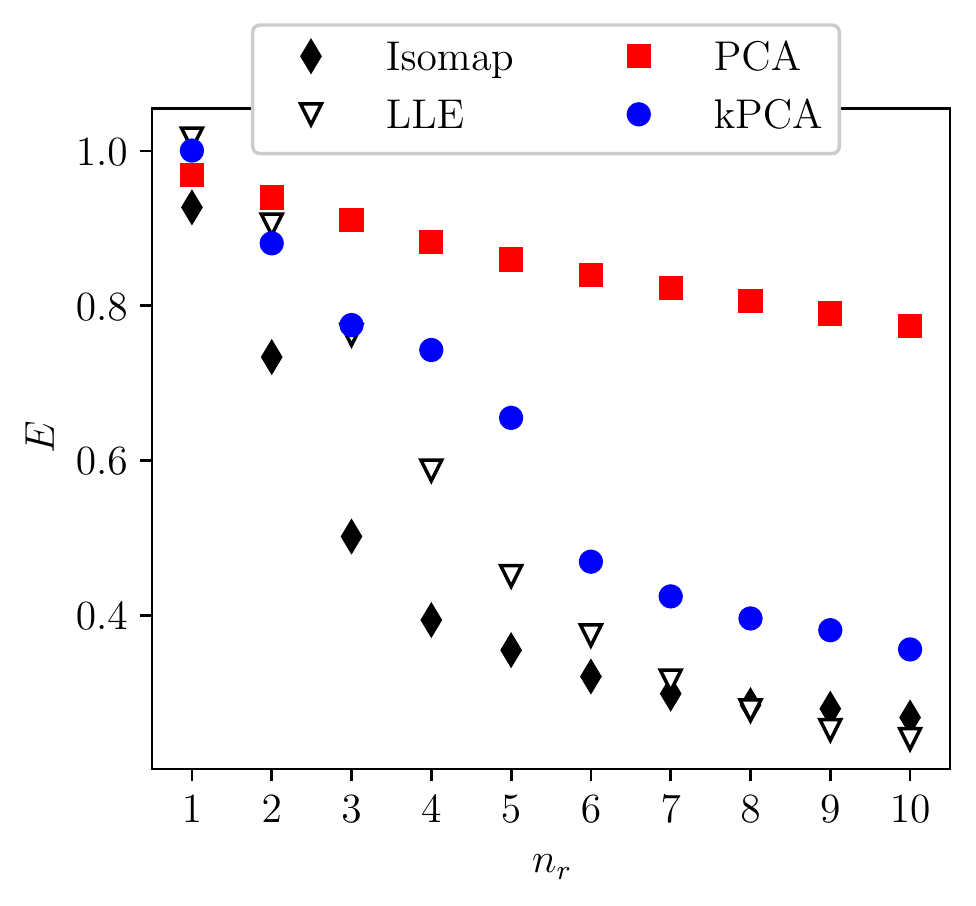}
		}
	\end{subfigure}
	\begin{subfigure}[\label{R_V_JET}]{
			\includegraphics[width=.45\textwidth ]{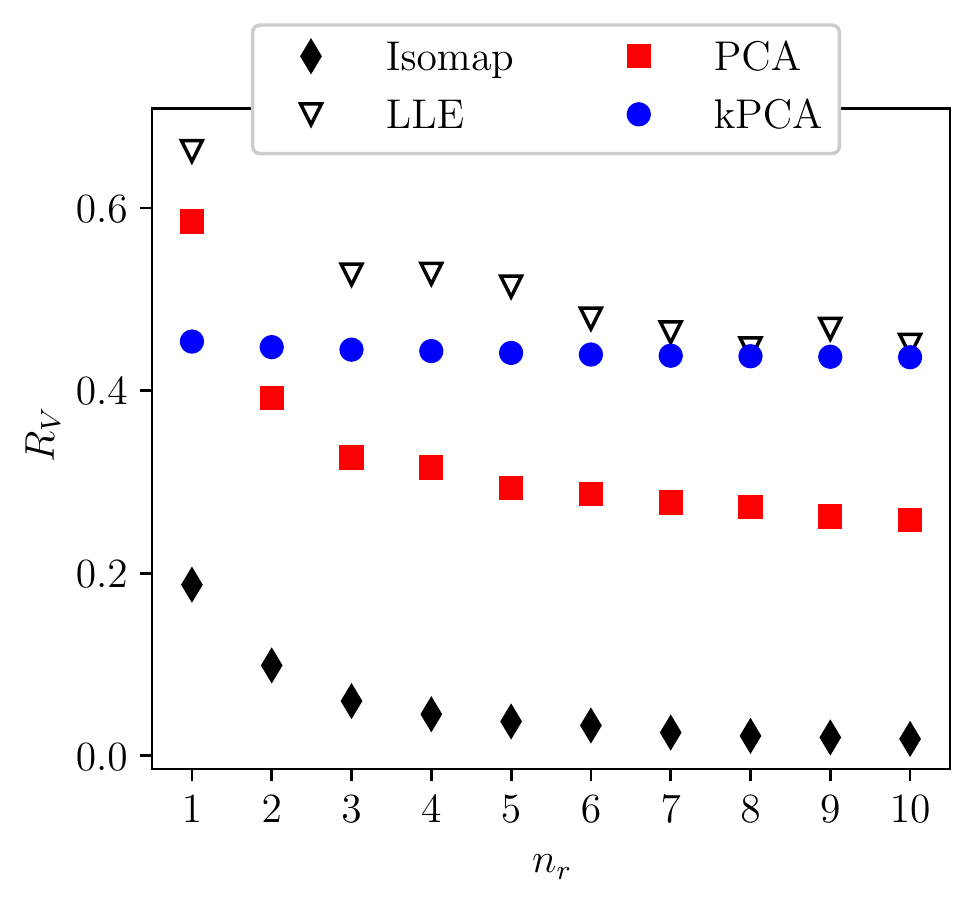}
		}
	\end{subfigure}	
	\caption{Same as figure \ref{Convergence_PIV_IM} and \ref{Convergence_CYL} but for the TR-PIV fields of impinging jet flow (Test Case 3, Section \ref{sec4p3}).}
	\label{Convergence_JET}
\end{figure}

We finally consider the flow field of an impinging gas jet (Sec. \ref{sec4p3}). Since the dataset is statistically stationary, the mean flow is subtracted prior to the dimensionality reduction and normalized by the largest velocity in the dataset. The same hyperparameters used in the previous test case are used for all methods. Figure \ref{Convergence_JET} shows the convergence of $E_B(n_r)$ and $R_V(n_r)$.

The multi-scale nature of this dataset makes constructing low-dimensional models considerably more difficult than in the previous test case. This is clearly visible from the slow convergence of the $l_2$ error in the PCA. Nevertheless, all nonlinear decompositions significantly outperform the PCA convergence, with ISOMAps giving the best convergence. As expected, ISOMAPs yield the best convergence also in terms of residual variance. Both kPCA and LLE perform well in $l_2$ error convergence but poorly in residual variance convergence. This confirms again that these metrics are generally not related.

\begin{figure}[ht]
			\centering
			\includegraphics[width=.6\textwidth ]{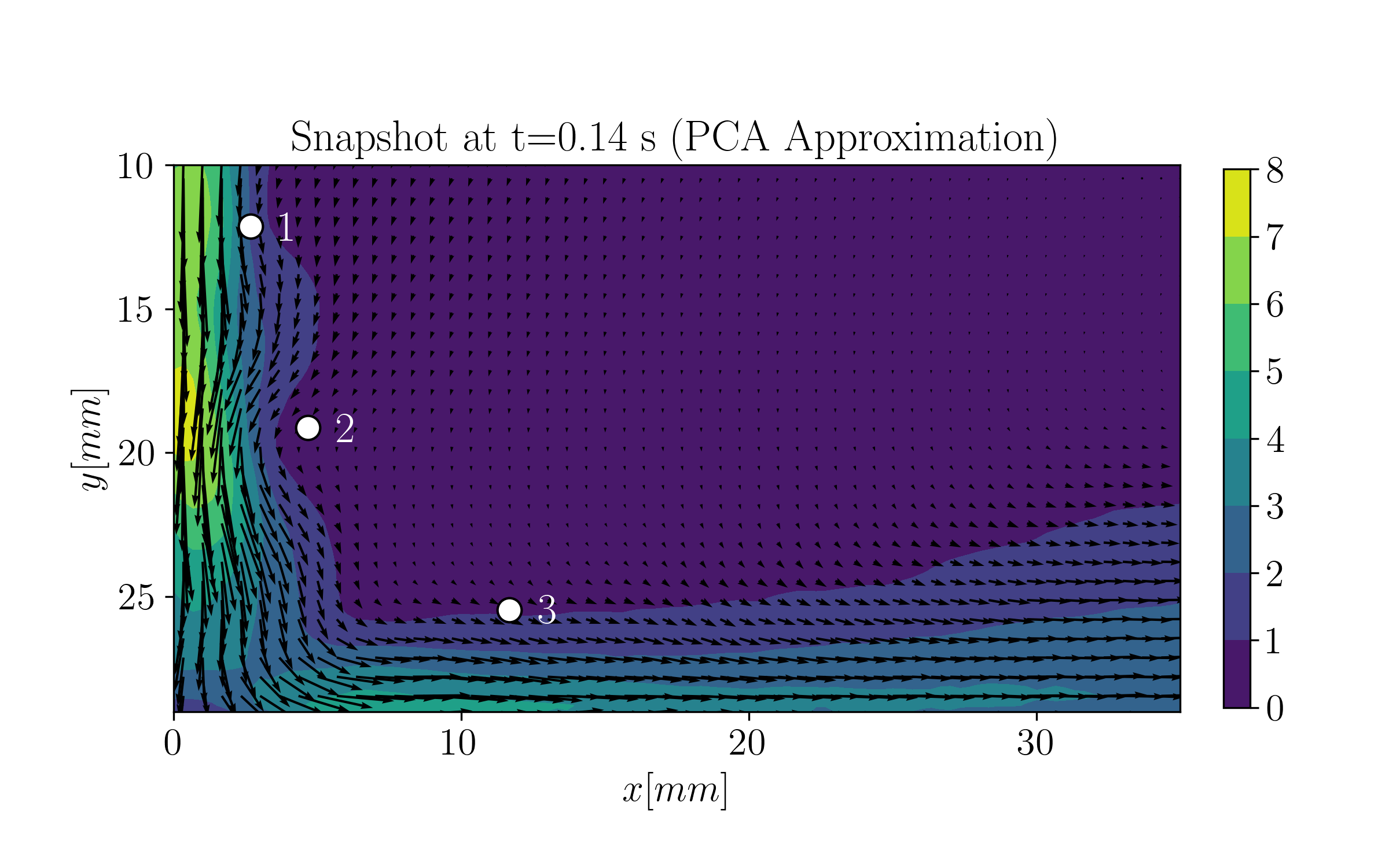}\\			
			\includegraphics[width=.3\textwidth ]{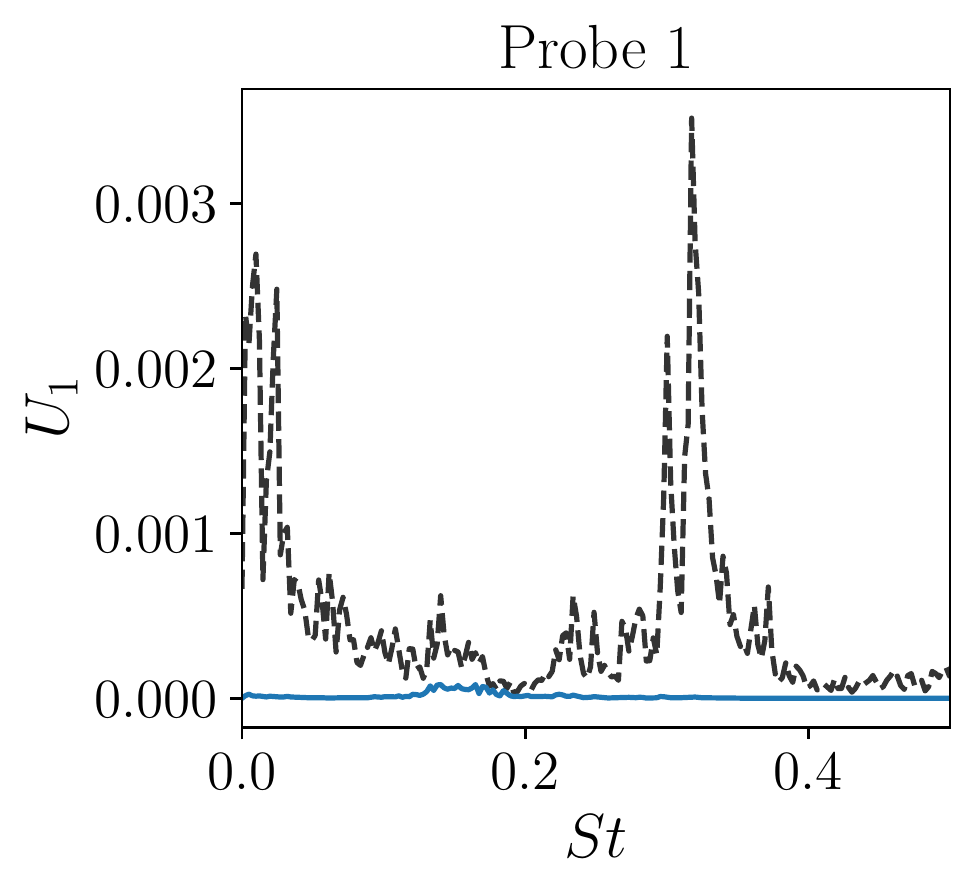}
			\includegraphics[width=.3\textwidth ]{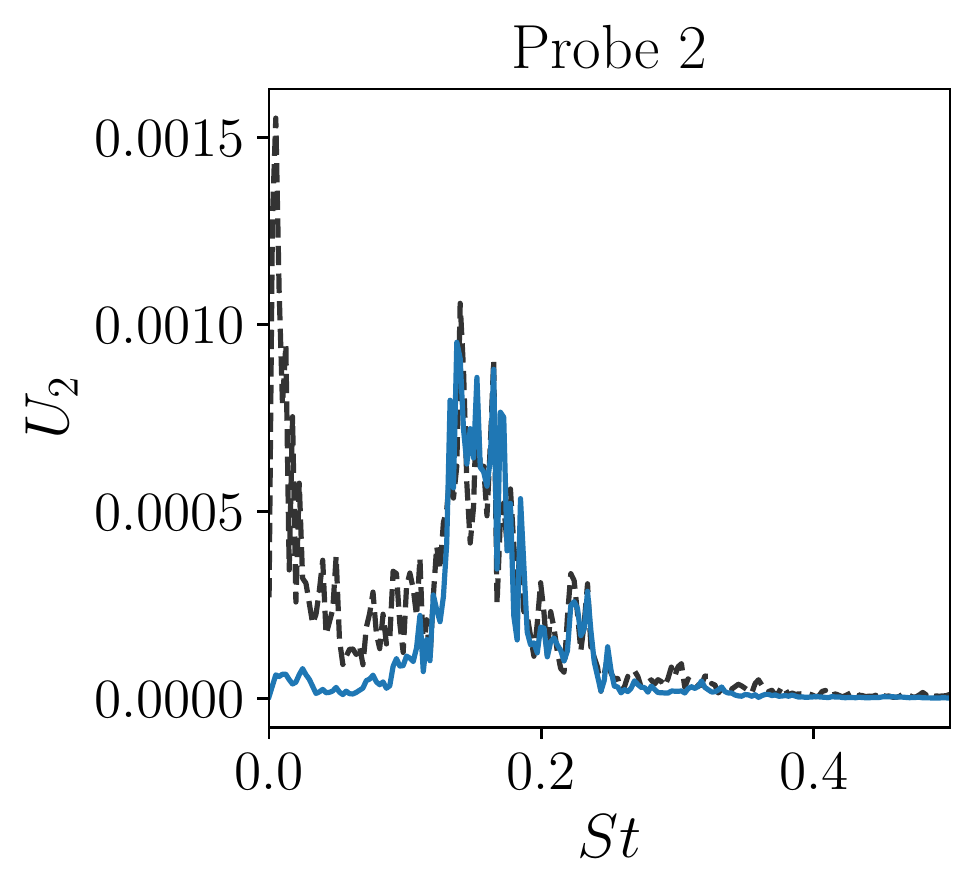}
			\includegraphics[width=.3\textwidth ]{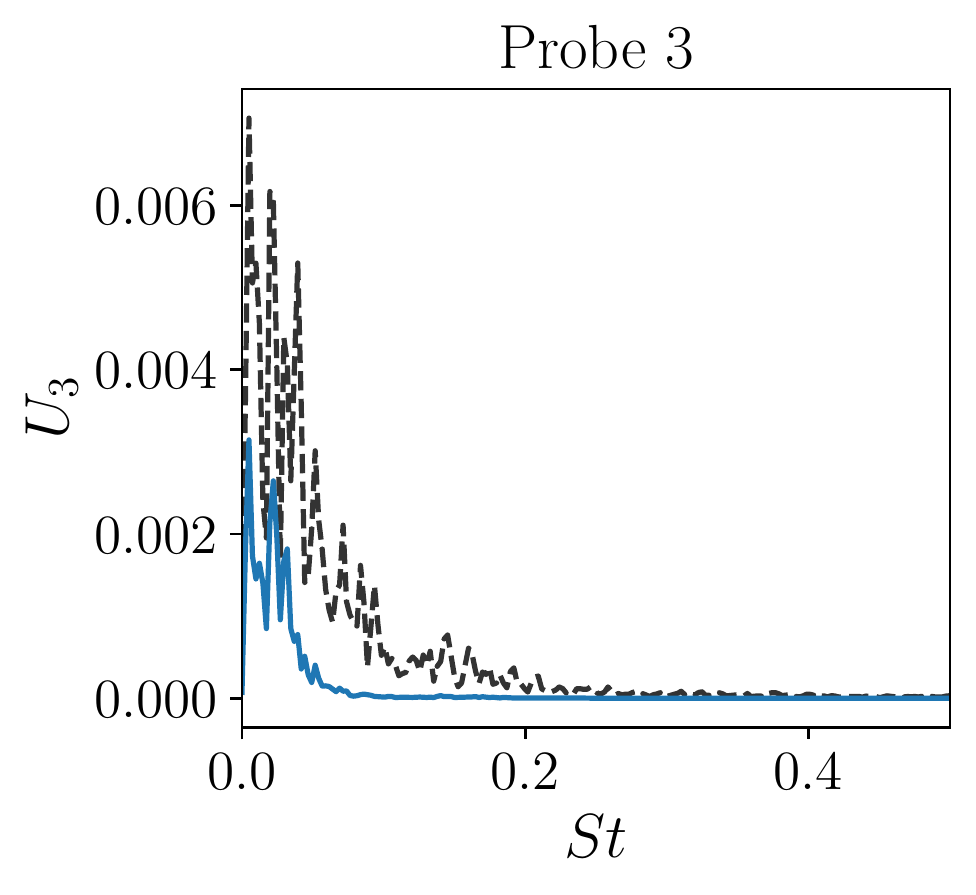}			
			\par\noindent\rule{\textwidth}{0.4pt}
			
 	        \includegraphics[width=.6\textwidth ]{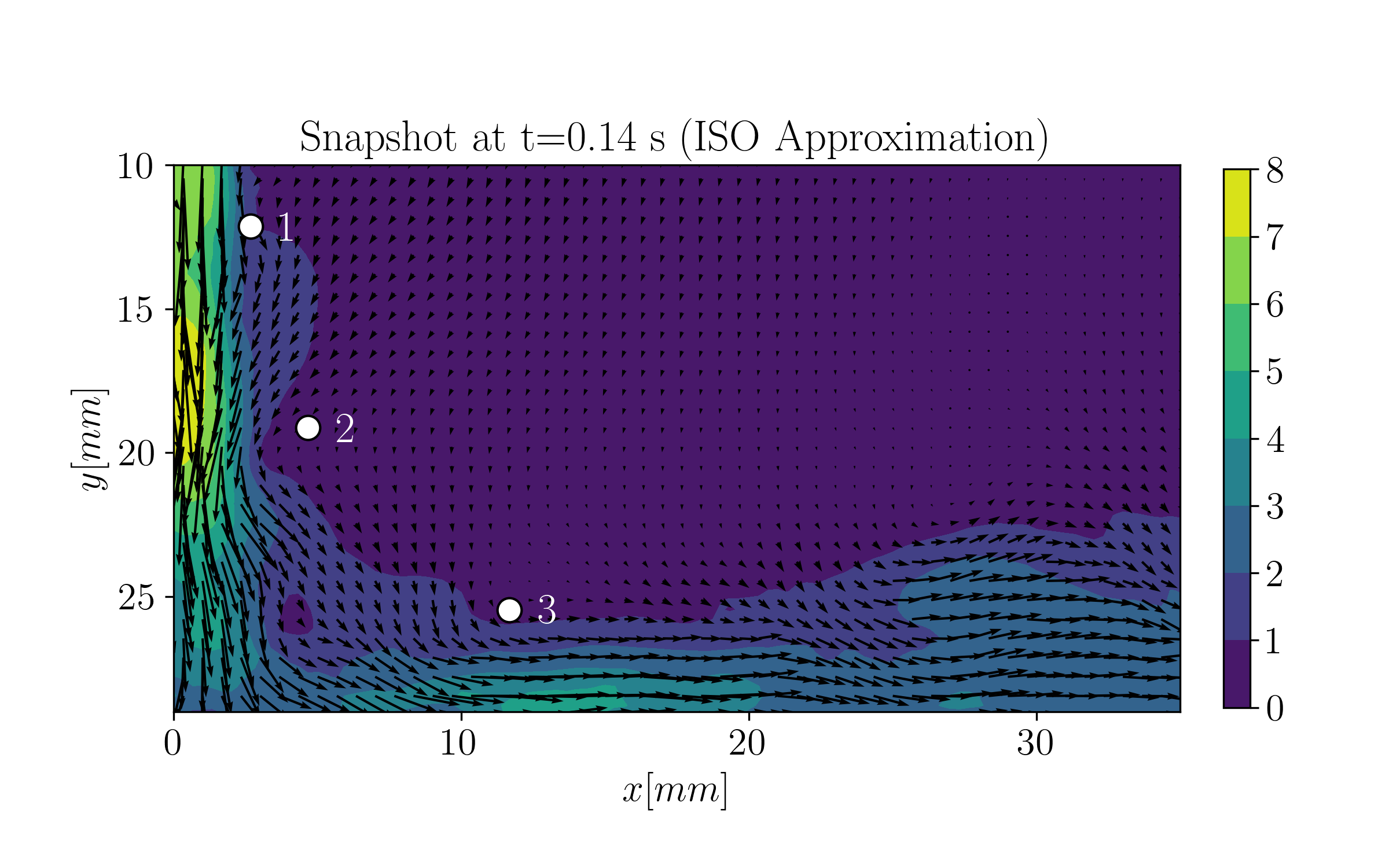}\\					
            \includegraphics[width=.3\textwidth ]{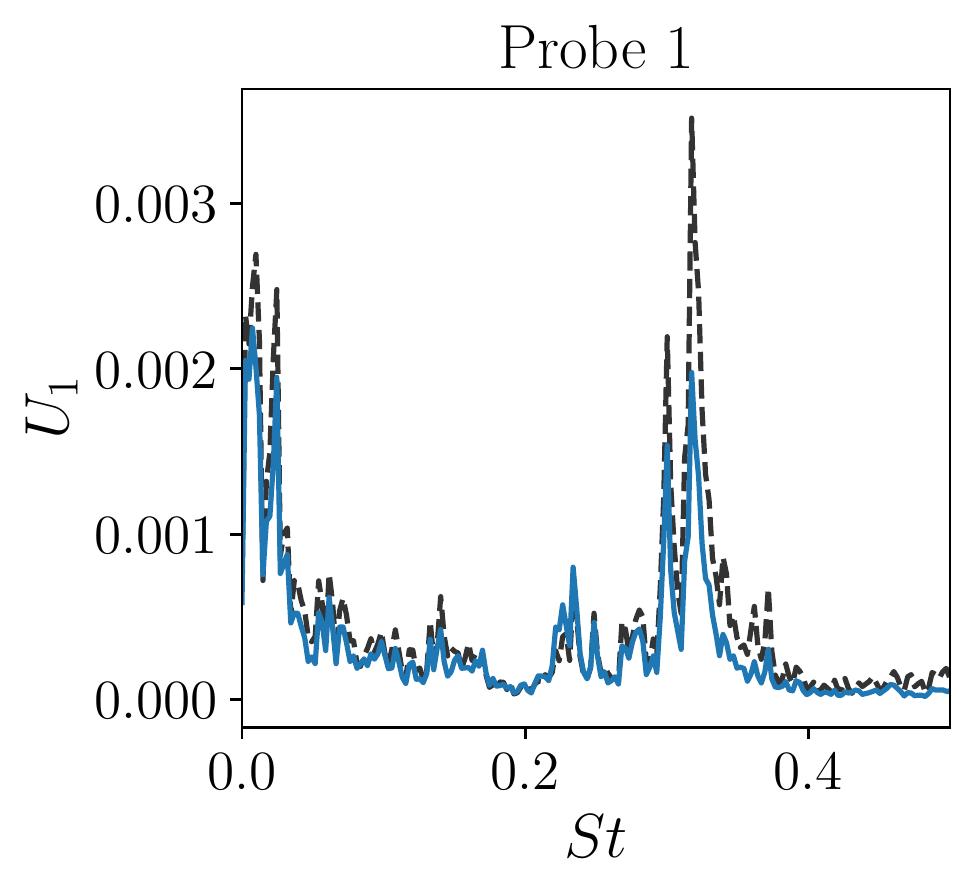}
            \includegraphics[width=.3\textwidth ]{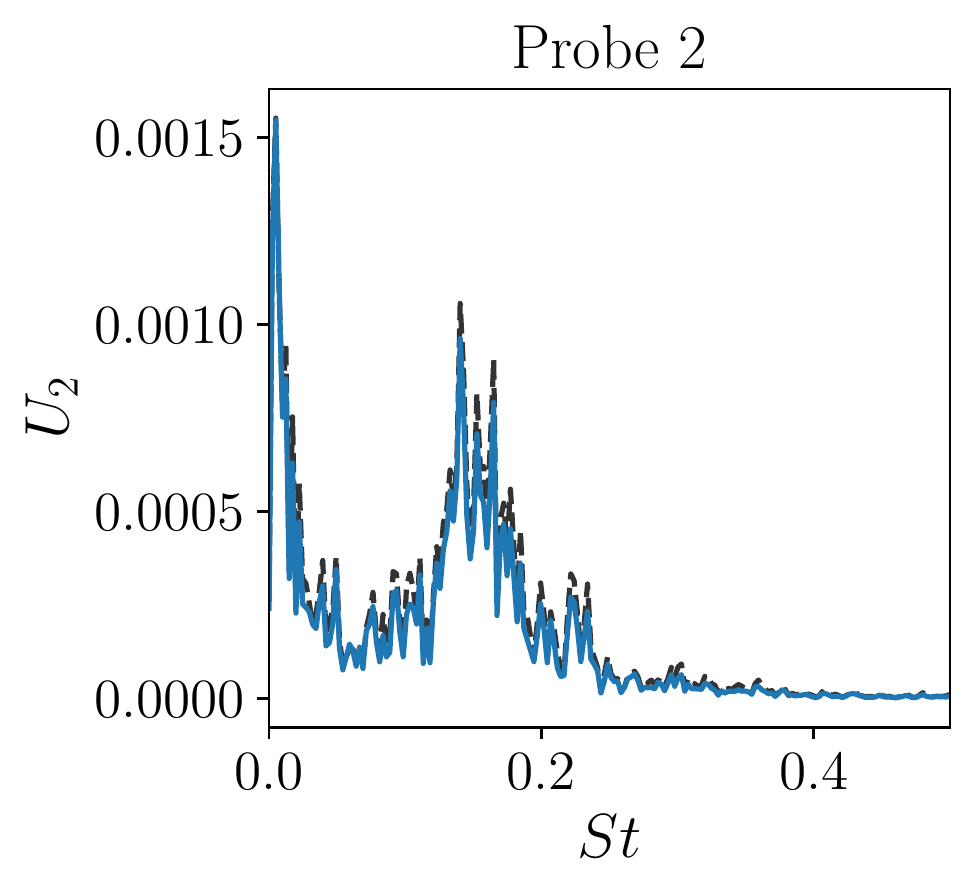}					
            \includegraphics[width=.3\textwidth ]{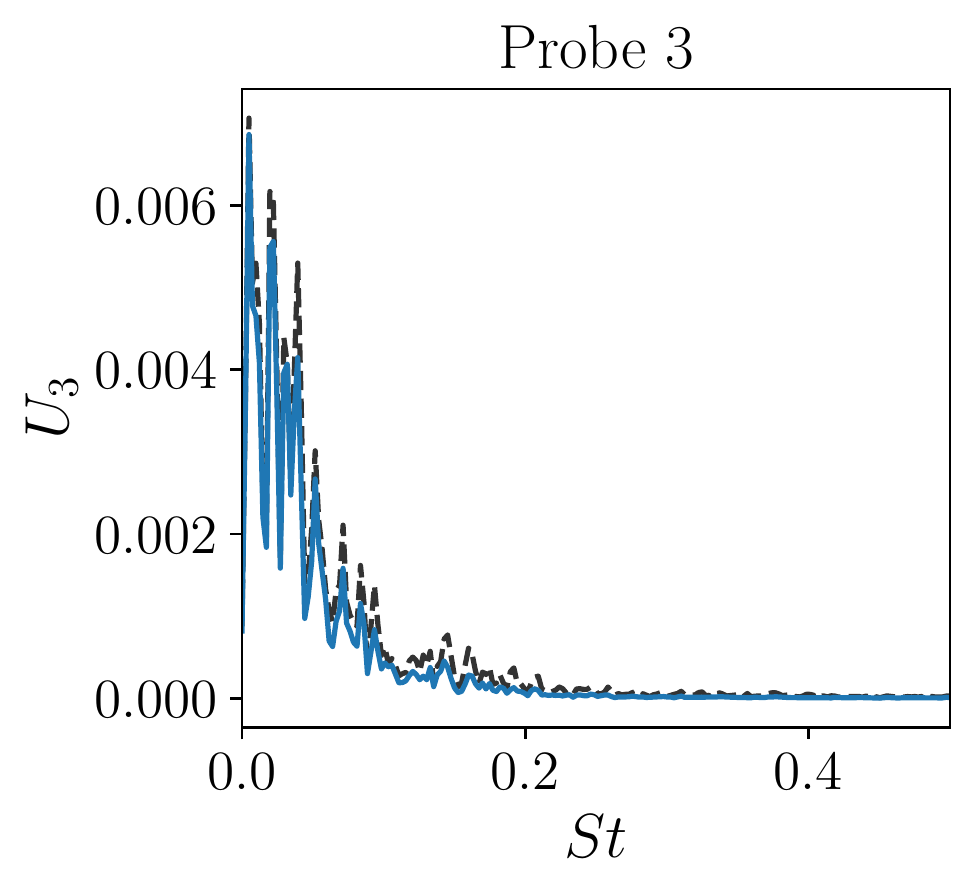}	
			\caption{Reconstruction of the TR-PIV measurements for the third test case (Section \ref{sec4p3}) with $n_r=4$ using PCA (top pannel) and ISOMAPs (bottom pannel). These figures should be confronted with Figure \ref{Original_Jet_RECO}, which shows the original snapshot. The plots under the velocity fields show the power spectral density of the velocity magnitude in the three probes (marked in the snapshots). The blue continuous lines refer to the reconstructed fields; black dashed lines refer to the original field.  } 
			\label{PCA_ISO_RECO}
		\end{figure}

Finally, the significant gains in the $l_2$ convergence are showcased in the reconstruction of a selected snapshot, considering $n_r=4$. This is the velocity field shown in Figure \ref{Original_Jet_RECO}. Figure \ref{PCA_ISO_RECO} shows the reconstruction for the PCA on the top and the reconstruction for the ISOMAps on the bottom.
In both cases, the figures are complemented with the power spectral densities of the velocity field in the three probes indicated in the figures (see also Section \ref{sec4p3}). The continuous blue line is used for the spectra in the reduced model, while the dashed black line is the corresponding spectra in the original field.

At $n_r=4$, the POD modes mostly capture the vortex shedding downstream the nozzle, in the shear layer region, at a distance of about 5 to 10 mm from the wall. Consequently, the PCA reconstruction captures most of the power spectral density in probe 2. On the other hand, it misses almost completely the flow features closer to the nozzle exit (where probe 1 is located) and offers a poor reconstruction of the wall jet region (where probe 3 is located). As a result, the flow field reconstructed via PCA appears much smoother than the original one, as only the large-scale features are retained. On the other hand, the ISOMAPs reconstruction recovers most of the features, as expected from the remarkable convergence in Figure \ref{Convergence_JET}, and as demonstrated by the snapshot and the power spectral density plots.
		
\section{Conclusions and Perspectives}\label{sec:6}

This work offered a concise introduction to nonlinear dimensionality techniques and the notions of autoencoders and manifold learning.
Autoencoders seek to preserve as much information as possible, while manifold learning techniques seek to preserve some measure of similarity. A general mathematical framework is reviewed and used to unify the treatment of kernel Principal Component Analysis (kPCA), isometric feature learning (ISOMAPs) and Locally Linear Embedding (LLE), as well as their relation to Principal Component Analysis (PCA), known as Proper Orthogonal Decomposition (POD) in fluid dynamics. Using a k-nearest linear interpolation as decoder, these decompositions were tested in three classic data processing problems for fluid dynamics: filtering, oscillatory pattern analysis and data compression.

In the selected filtering problem, namely the background noise removal in PIV images, it was found that none of the implemented nonlinear tools outperforms the PCA/POD. While we refrain from drawing definite conclusions from a single test case, this result highlights that the idea of preserving similarity, whether globally (as in ISOMAPs) or locally (as in LLE), might not be relevant to a filtering process. 

In the oscillatory pattern detection problem, namely the dimensionality reduction of a transient cylinder wake flow, it was found that ISOMAPs and PCA have comparable $l_2$ reconstruction error if the reduced dimension is $n_r=3$, as it is customarily done for this kind of flows. The resulting manifolds are qualitatively similar. Both are surfaces of revolution, but the generating function is approximately linear in the PCA (hence generating a conical frustum) and approximately sigmoidal in the ISOMAPs. The ISOMAPs outperform the PCA in the residual variance convergence; thus, one might conclude that the identified manifold is closer to the `true' underlying one. The performances of kPCA and LLE are comparable (they outperform the PCA in residual variance but not in the reconstruction error) and result in similar manifolds: a conical frustum with opposite orientation to the ones in PCA and ISOMAPs. This shows that a `small' difference in the implemented metrics can lead to significant differences in the system trajectory in the reduced space. One should keep this in mind when analyzing the topology of low-dimensional manifolds derived from data.

Finally, in the investigated data compression problem, which considered the flow of an impinging gas jet, all nonlinear methods significantly outperformed the PCA (POD) in terms of $l_2$ error convergence. This result is remarkable considering the widely different scales characterizing this test case and the poor convergence of the PCA. The ISOMAPs and LLE bring the convergence error below $40\%$ at $n_r=10$. Inspecting the reconstruction of the flow field and the power spectral density in three selected probes showed that the ISOMAPs preserve most of the relevant features even at $n_r=4$. 

As a general conclusion, it is worth highlighting that all the analyzed test cases show that the minimization of the $l_2$ reconstruction error and the residual variance are vastly different (if not contrasting) objectives, and methods that excel in one metric might fail in the other. Nevertheless, the ISOMAPs show the best overall performance among the implemented methods. 
In the author's opinion, whether this method will reach the ubiquity of PCA/POD depends on how much we can adapt its formalism to (and our interpretation of) classic problems.

In a filtering problem, for example, the better convergence of ISOMAPs implies that this is also more prone to retain noise in the autoencoded result. A multi-scale extension of the problem, whereby the ISOMAPs is used at different scales, could lead to a powerful filtering approach. In the pattern recognition problem, reduced representations from ISOMAPs or kPCA can reveal essential dynamics, but their mapping to the space domain is conceptually much more involved than the simple projection in a linear approach such as PCA. 

The nonlinearity of these methods brings an important conceptual shift, and the familiar notions of \emph{modes}, \emph{orthogonality} and \emph{Galerkin projection} are no longer relevant. Besides complicating the spatial localization of the essential dynamics, this significantly challenges the construction of `intrusive' low dimensional models, i.e. the integration of low order representation from data with the governing equations. Finally, an important line of research is the decoding problem, with particular emphasis on the interpolative nature of methods such as the k-nearest linear approach used in this work or the regression techniques like the kernel Ridge regression encountered in the literature of kPCA. While this work focused on `in-sample' error in the reconstruction (i.e. decoding the same data that was used to train the encoder), the robustness of the `out-of-sample' reconstruction (i.e. decoding new data) remains to be assessed in problems of interest for fluid dynamics. 

These exciting problems will surely attract and challenge many enthusiastic fluid dynamicists to continue erasing the boundaries between ordinary data processing and machine learning.

\clearpage
		
	\bibliography{Mendez_MST_2022.bib}

	\end{document}